\documentclass[aps,tightenlines,nofootinbib,prd]{revtex4}
\usepackage{amsmath,amscd,amssymb}
\usepackage{color,epsfig}
\usepackage{xfrac}
\usepackage{latexsym}
\usepackage{mathrsfs}
\usepackage{ulem}

\newcommand{\dslash}{\partial \hskip -0.5em /}
\newcommand{\vslash}{v \hskip -0.5em /}
\newcommand{\aslash}{a \hskip -0.5em /}
\newcommand{\nslash}{n \hskip -0.5em /}

\newcommand{\Vek}[1]{{\boldsymbol#1}}

\newcommand{\imu}{{\rm i}}

\begin{document}

\title{Nucleon Structure Functions from the NJL-Model Chiral Soliton}

\author{I. Takyi, H. Weigel}

\affiliation{
Institute for Theoretical Physics, Physics Department, Stellenbosch University,
Matieland 7602, South Africa}

\begin{abstract}
We present numerical simulations for unpolarized and polarized structure functions
in a chiral soliton model. The soliton is constructed self-consistently from
quark fields from which the structure functions are extracted. Central to the 
project is the implementation of regularizing the Dirac sea (or vacuum) contribution 
to structure functions from first principles. We discuss in detail how sum rules
are realized at the level of the quark wave-functions in momentum space. 
The comparison with experimental data is convincing for the polarized structure 
functions but exhibits some discrepancies in the unpolarized case. The vacuum
contribution to the polarized structure functions is particularly small.
\end{abstract}

\maketitle

\section{Introduction}

Perhaps the most convincing evidence for the quark substructure of baryons emerges
from Deep Inelastic Scattering (DIS). The conjuction of perturbative Quantum Chromo
Dynamics (QCD) and the parton model successfully explains the wealth of DIS data collected 
over the past decades \cite{Mu87,Ro90,Yn93}. However, these are not fully first principle 
calculations as the hadron wave-functions cannot (yet) be directly computed in QCD. 
Rather, in the spirit of the parton model, quark distributions are parameterized 
and subjected to perturbative QCD analysis \cite{Pumplin:2002vw}. On the other hand
there are many phenomenological models based on various aspects of QCD that (attempt to)
describe (static) properties of hadrons with particular focus on the nucleon. Examples
are the non-relativistic quark model \cite{Ko79}, the MIT bag model \cite{Thomas:1982kv}
relativistic quark-diquark models \cite{Maris:2003vk}, chiral soliton models 
\cite{Weigel:2008zz}; just to name a few. In principle any of these approaches 
should also be capable to predict nucleon structure functions. This is particular 
challenging for chiral soliton models that are formulated as bosonized action functionals 
hiding the quark substructure of hadrons. In this respect the Nambu-Jona-Lasino (NJL)
or chiral quark soliton model \cite{Alkofer:1994ph,Christov:1995vm} is special: based
on a quark self-interaction the bosonization process can be traced step by step.

In the past nucleon structure functions have indeed been computed from the chiral quark 
soliton model and mainly two approaches were followed. Within the {\it valence quark only 
approximation} \cite{Weigel:1996kw,Weigel:1996jh} the observation that though the Dirac sea 
(or vacuum) is essential to form the soliton, the vacuum contribution to nucleon properties 
is only moderate thereby justifying neglecting its contribution to the structure functions.
Though this approximation has empirical support, it is formally incomplete and does not follow 
from a systematic expansion scheme. In parallel, studies on the quark distributions were performed 
\cite{Diakonov:1996sr,Diakonov:1997vc,Pobylitsa:1998tk,Wakamatsu:1997en,Wakamatsu:1998rx,Wakamatsu:2003wg,Wakamatsu:2002kq}
that identified the model quark degrees of freedom with those using the operator product expansion 
in the analysis of DIS. These studies included vacuum contributions. These are plagued 
by ultra-violet divergences requiring an {\it a posteriori} implementation of regularization
by a single Pauli-Villars subtraction imposed onto the distributions.  Unfortunately, this 
is not stringent since there are terms in the action that do not undergo 
regularization, {\it e.g.} to maintain the axial anomaly as measured by the decay of 
the neutral pion into two photons. Generally, the model has quadratic divergences (most notably
the gap equation) and a single subtraction may or may 
not be sufficient to remove all divergences. In the present project we therefore 
include the vacuum contributions to the nucleon structure functions (i) without identifying 
the NJL model quarks field with those of QCD, and (ii) by implementing the mandatory regularization 
already at the level of the defining action functional. The first issue is addressed by noting 
that the model emulates the chiral symmetry of QCD and thus produces the same symmetry currents,
in particular the electromagnetic one. To address the second issue we recall that DIS is 
described by the hadronic tensor $W_{\mu\nu}$ which is the Fourier transform of the nucleon 
matrix element of the commutator of two electromagnetic current operators. Though there is 
no direct implementation of this commutator when bosonizing the NJL model, we take advantage
of its relation to the Compton tensor $T_{\mu\nu}$ which itself is computed from the time-ordered 
product of these currents. Time-ordered products are straightforwardly included in a path integral
formalism within which bosonization is conducted. Regularizing this path integral by multiple 
Pauli-Villars subtractions proves most appropriate because it allows to trace the quark that 
carries the large momentum in the Bjorken limit. This formalism was developed already some
time ago \cite{Weigel:1999pc} but its numerical simulation has been long outstanding. It will
be central to the study presented here.

This paper is organized as follows: In Section \ref{sec:model} we introduce the NJL model with
emphasis on describing the regularization procedure in Minkowski space. The formulation in 
Minkowski space is advantageous to identify the absorptive part of $T_{\mu\nu}$ which leads to 
$W_{\mu\nu}$. In Section \ref{sec:SFinNJL} we review the formalism from Ref. \cite{Weigel:1999pc} 
of how to obtain the hadronic tensor in this model and in particular the role of the Bjorken 
limit. The NJL soliton description is explained in Section \ref{sec:sol}. We discuss the formalism 
to obtain the structure functions via the hadronic tensor from quark spinors that self-consistently
interact with the soliton in Section \ref{sol_st_fct}. Subsequently (Section \ref{sec:SR})
we describe the way in which this formalism builds up the sum rules. Numerical results
are presented in Section \ref{sec:NR}. This analysis also includes the perturbative 
evolution to the scale at which experimental data are taken. As stressed above,
the model structure functions are identified from symmetry currents, not by 
equating QCD degrees of freedom. However, the evolution makes this unavoidable for 
the lack of any sensible alternative.  We briefly conclude and summarize in 
Section \ref{sec:concl}. Finally we leave technical details to four Appendixes.

Some preliminary results extracted from this paper have been put forward in
Ref. \cite{TakyiQNP18}

\section{The model}
\label{sec:model}

We formulate the regularized action of the bosonized Nambu-Jona-Lasino (NJL) model
in Minkowski space as the sum of three pieces
\begin{align}
\mathcal{A}_{\rm NJL}&=\mathcal{A}_{\rm R}+\mathcal{A}_{\rm I}
+\frac{1}{4G}\int d^4x\,
{\rm tr}\left[S^2+P^2+2m_0S\right]
\label{act1} \\
\mathcal{A}_{\rm R}&=-i\frac{N_C}{2}
\sum_{i=0}^2 c_i {\rm Tr}\, {\rm log}
\left[- \mathbf{D} \mathbf{D}_5 +\Lambda_i^2-i\epsilon\right]\,,\cr
\mathcal{A}_{\rm I}&=-i\frac{N_C}{2}
{\rm Tr}\, {\rm log}
\left[-\mathbf{D} \left(\mathbf{D}_5\right)^{-1}-i\epsilon\right]\, .
\nonumber\end{align}
The subscripts $R$ and $I$ for real and imaginary refer to the respective properties 
after Wick-rotation to Euclidean space. In Minkowski space this corresponds to the 
use of two distinct Dirac operators
\begin{align}
\mathbf{D}&=\imu\dslash-\left(S+\imu\gamma_5P\right) +\vslash +\aslash\gamma_5 
=:\mathbf{D}^{(\pi)}+\vslash +\aslash\gamma_5 \cr
\mathbf{D}_5&=-\imu\dslash-\left(S+\imu\gamma_5P\right) -\vslash +\aslash\gamma_5 
=:\mathbf{D}_5^{(\pi)}+\vslash +\aslash\gamma_5\,.
\label{DirOp}\end{align}
Here, as in the local part of Eq.~(\ref{act1}), $S$ and $P$ are scalar and pseudoscalar 
fields that represent physical particles. Furthermore $v_\mu$ and $a_\mu$ are source fields. 
When expanding the action with respect to these sources the linear terms couple to the 
vector and axial vector currents. Since $\mathcal{A}_{\rm I}$ is (conditionally) finite 
in the ultra-violet, only $\mathcal{A}_{\rm R}$ undergoes regularization. Its dynamical 
content is quadratically divergent\footnote{The cosmological constant, computed from 
setting all fields to their vacuum expectation values, is quartically divergent and must
be subtracted when contributing.} and we thus require two subtractions. In the 
Pauli-Villars scheme they are implemented as
\begin{equation}
c_0=1\, ,\quad \Lambda_0=0\, ,\quad \sum_{i=0}^2c_i=0
\quad {\rm and}\quad \sum_{i=0}^2c_i\Lambda_i^2=0\,.
\label{pvcond}\end{equation}
In practice we will reduce the number of regularization parameters by assuming 
$\Lambda_1  \to \Lambda_2 = \Lambda $ which translates into the general prescription
$\sum_i c_i f(\Lambda_i^2)= f(0)-f(\Lambda^2)+\Lambda^2 f' (\Lambda^2 )$. Nevertheless
we always write the regularization as in Eq.~(\ref{act1}).

The variation of the action with respect to the scalar field $S$ yields the gap equation
\begin{equation}
\frac{1}{2G}\left(m-m_0\right)
=-4iN_C m\sum_{i=0}^2c_i
\int\frac{d^4k}{(2\pi)^4}
\left[-k^2+m^2+\Lambda_i^2-i\epsilon\right]^{-1}
\label{gap} \end{equation}
that determines the vacuum expectation value of the scalar field $\langle S\rangle =m$.
Replacing $S$ by $\langle S\rangle$ in $\mathbf{D}$ shows that $m$ is the fermion mass and
is thus called the {\it constituent quark mass}. The chiral field $U$ defines the non-linear 
representation for the isovector pion field $\Vek{\pi}$ 
\begin{equation}
S+\imu P=m U =m\,{\rm exp}\left[\imu \frac{g}{m}\Vek{\pi}\cdot\Vek{\tau}\right]\,.
\label{xcircle}\end{equation}
We obtain the pion propagator by expanding $\mathcal{A}$ to quadratic order and extract
the pion mass from the pole
\begin{equation}
m_\pi^2=\frac{1}{2G}\frac{m_0}{m}\frac{1}{2N_C\Pi(m_\pi^2)}
\label{mpi} \end{equation}
and requiring a unit residuum at the pole fixes the quark-pion coupling
\begin{equation}
\frac{1}{g^2}=4N_C \frac{d}{dq^2}
\left[q^2 \Pi(q^2)\right]\Bigg|_{q^2=m_\pi^2} \, .
\label{gpcoup} \end{equation}
The above polarization function is
\begin{equation}
\Pi(q^2)=\int_0^1 dx\, \Pi(q^2,x) \qquad {\rm with}\qquad
\Pi(q^2,x)=-i\sum_{i=0}^2 c_i\,
\frac{d^4k}{(2\pi)^4}\,
\left[-k^2-x(1-x)q^2+m^2+\Lambda_i^2-i\epsilon\right]^{-2}\,.
\label{specfct} \end{equation}
Finally we get the pion decay constant $f_\pi$ from the coupling to the axial current.
To this end we expand $\mathcal{A}$ to linear order in both $\Vek{\pi}$ and $a_\mu$.
The result is $f_\pi=4N_C m g \Pi(m_\pi^2)$. Using the empirical data $m_\pi=138{\rm MeV}$
and $f_\pi=93{\rm MeV}$ fixes two of the three ($m_0$, $G$ and $\Lambda$) model constants.
It is customary to use the constituent quark mass $m$ as the single tunable parameter.

\section{Structure functions in the NJL model}
\label{sec:SFinNJL}

The hadronic tensor of DIS is the matrix element of the commutator 
of electro-magnetic current operators
\begin{equation}
W_{\mu \nu}(q;H) = \frac{1}{4\pi}
\int d^4x\, {\rm e}^{\imu q\cdot \xi}\,
\Big\langle H \Big| [ J_{\mu}(\xi),J_{\nu}^{\dagger}(0)]
\Big| H \Big\rangle\, ,
\label{hten1}\end{equation}
where $q$ is the momentum of the virtual photon and $H$ refers to either the pion or 
the nucleon target\footnote{The hadronic tensor can be written as the matrix element of the 
commutator for the lowest energy hadron in a given baryon number sector.}. 
This tensor is decomposed into Lorentz structures whose coefficients are form factors that
turn into the structure functions in the so-called Bjorken limit. Labeling the 
target momentum by $p$ this limit is defined as
\begin{equation}
Q^2=-q^2\to\infty  \qquad {\rm with}\quad
x=\frac{Q^2}{2p\cdot q}\quad {\rm fixed}\,.
\label{eq:defBlim}\end{equation}
Often $x$ is referred to as the Bjorken variable.

By the optical theorem, $W_{\mu \nu}$ is proportional to the absorptive part of the 
Compton amplitude
\begin{equation}
W_{\mu \nu}(q;H)=\frac{1}{2\pi}\, {\rm Im}\, T_{\mu \nu}(q;H)\, .
\label{Comp1}\end{equation}
The latter is the  matrix element of a time-ordered
product of the currents
\begin{equation}
T_{\mu \nu}(q;H) = \imu
\int d^4\xi {\rm e}^{\imu q\cdot \xi}\,
\Big\langle H\Big| T\left\{J_{\mu}(\xi)
J_{\nu}^{\dagger}(0)\right\}\Big| H \Big\rangle\,.
\label{Comp2}\end{equation}
With this relation we implement regularization from first principles because the time-ordered
product is immediately obtained from the action by functional differentiation
\begin{equation}
T\left\{J_\mu(\xi) J_\nu(0)\right\}=
\frac{\delta^2}{\delta v^\mu(\xi)\delta v^\nu(0)}
\mathcal{A}_{\rm NJL}(v)\Bigg|_{v_\mu=0}\,,
\label{disp2}\end{equation}
where $v_\mu$ is the photon field introduced by minimal coupling. Within the NJL model the 
photon couples to the quarks inside the hadron. As discussed comprehensively in 
Ref.~\cite{Weigel:1999pc} the evaluation of $T_{\mu\nu}$ becomes feasible in the Bjorken 
limit. Then the quark propagator that carries the large photon momentum can be 
identified and thus be taken to be that of a free massless fermion. Thus the functional 
derivative from Eq.~(\ref{disp2}) when applied to the real part simplifies to 
differentiating
\begin{equation}
\mathcal{A}_{\Lambda,{\rm R}}^{(2,v)}=
-i\frac{N_C}{4}\sum_{i=0}^2c_i
{\rm Tr}\,\left\{\left(-\mathbf{D}^{(\pi)}\mathbf{D}^{(\pi)}_5+\Lambda_i^2\right)^{-1}
\left[{\cal Q}^2\vslash\left(\dslash\right)^{-1}\vslash\mathbf{D}^{(\pi)}_5
-\mathbf{D}^{(\pi)}(\vslash\left(\dslash\right)^{-1}\vslash)_5
{\cal Q}^2\right]\right\}.
\label{simple6}\end{equation}
At this point it is important to explain the crucial role of the subscript `5' 
attached to the second term in square brackets of Eq.~(\ref{simple6}). For this 
second term we have to recall that the (inverse) derivative operator in
$\vslash\left(\dslash\right)^{-1}\vslash$ is actually
associated with the expansion of $\mathbf{D}_5$. When comparing this
$\gamma_5$--odd operator to the ordinary Dirac operator in Eq.~(\ref{DirOp})
one observes immediately that $\mathbf{D}_5$ has a relative sign between the 
derivative operator $i\dslash$ and the axial source $\aslash\gamma_5$. 
Therefore the axial--vector component of
$(\vslash\left(\dslash\right)^{-1}\vslash)_5$
requires a relative sign.
With $S_{\mu\rho\nu\sigma}=g_{\mu\rho}g_{\nu\sigma}
+g_{\rho\nu}g_{\mu\sigma}-g_{\mu\nu}g_{\rho\sigma}$, that is
\begin{equation}
\gamma_\mu\gamma_\rho\gamma_\nu
=S_{\mu\rho\nu\sigma}\gamma^\sigma
-i\epsilon_{\mu\rho\nu\sigma}\gamma^\sigma\gamma^5
\quad {\rm while} \quad
(\gamma_\mu\gamma_\rho\gamma_\nu)_5
=S_{\mu\rho\nu\sigma}\gamma^\sigma+
i\epsilon_{\mu\rho\nu\sigma}\gamma^\sigma\gamma^5\, .
\label{defsign}\end{equation}
In Ref. \cite{Weigel:1999pc} this modification was formally shown to be
consistent with the affected sum rules. In Section \ref{sec:SR} we see on the level
of the momentum space quark wave-functions that the structure functions
computed on the basis of Eq.~(\ref{defsign}) indeed fulfill the sum rules.

Similarly, in the Bjorken limit, the imaginary part becomes
\begin{equation}
\mathcal{A}_{\Lambda,{\rm I}}^{(2,v)}=
-i\frac{N_C}{4}
{\rm Tr}\,\left\{\left(-\mathbf{D}^{(\pi)}\mathbf{D}^{(\pi)}\right)^{-1}
\left[{\cal Q}^2\vslash\left(\dslash\right)^{-1}\vslash\mathbf{D}^{(\pi)}_5
+\mathbf{D}^{(\pi)}(\vslash\left(\dslash\right)^{-1}\vslash)_5
{\cal Q}^2\right]\right\}\, .
\label{simple7}\end{equation}
These expression are still quite formal and we will use them to obtain nucleon 
structure functions in Section \ref{sol_st_fct}. We emphasize that these expressions
are directly deducted from the regularized action in Eq.~(\ref{act1}) and that 
no further assumption about the regularization has been made.

In Ref.~\cite{Weigel:1999pc} it has been shown that applying this formalism to the
pion relates its structure function to the spectral function from Eq.~(\ref{specfct})
as $F(x)=\frac{5}{9} (4N_C g^2) \frac{d}{dp^2}\left[p^2\Pi(p^2,x)\right]\Big|_{p^2=m_\pi^2}$, 
a result that was previously obtained from the analysis of the handbag diagram
in Refs.~\cite{Frederico:1994dx,Davidson:1994uv}.

\section{NJL model soliton}
\label{sec:sol}

We construct the soliton from static meson configurations by introducing
a Dirac Hamiltonian $h$ via
\begin{equation}
\imu\mathbf{D}^{(\pi)}=\beta(\imu\partial_t-h) \quad {\rm and}\quad
\imu\mathbf{D}^{(\pi)}_5=(-\imu\partial_t-h)\beta\, .
\label{defh}
\end{equation}
Its diagonalization
\begin{equation}
h\Psi_\alpha = \epsilon_\alpha \Psi_\alpha \, ,
\label{diagh}
\end{equation}
yields eigen-spinors $\Psi_\alpha=\sum_{\beta}V_{\alpha\beta}\Psi^{(0)}_\alpha$ 
($\Psi^{(0)}_\alpha$ are free Dirac spinors in a spherical basis, see Appendix 
\ref{app:sol}) and energy eigenvalues $\epsilon_\alpha$.  
The hedgehog configuration minimizes the action in the unit baryon number sector and 
introduces the chiral angle $\Theta(r)$ via
\begin{equation}
h=\Vek{\alpha}\cdot\Vek{p} +\beta\, m\, U_5(\Vek{r}) 
\qquad {\rm where}\qquad 
U_5(\Vek{r})={\rm exp} \left[\imu \hat{\Vek{r}}\cdot\Vek{\tau}\,
\gamma_5 \Theta(r)\right]\, .
\label{hedgehog}
\end{equation}
With the boundary conditions $\Theta(0)=-\pi$ and 
$\lim_{r\to\infty}\Theta(r)=0$ the diagonalization, Eq.~(\ref{diagh}) yields
a distinct, strongly bound level, $\Psi_{\rm v}$, referred to as the valence 
quark level \cite{Alkofer:1994ph}. Its (explicit) occupation ensures 
unit baryon number. The functional trace in $\mathcal{A}_R$ is computed 
as an integral over the time interval $T$ and a discrete sum over 
the basis levels defined by Eq.~(\ref{diagh}). In the limit $T\to\infty$ 
the vacuum contribution to the static energy is then extracted 
from $\mathcal{A}_R\to-TE_{\rm vac}$. Collecting pieces, we obtain
the total energy functional as \cite{Alkofer:1994ph,Christov:1995vm}
\begin{equation}
E_{\rm tot}[\Theta]=
\frac{N_C}{2}\left[1+{\rm sign}(\epsilon_{\rm v})\right]
\epsilon_{\rm v}
-\frac{N_C}{2}\sum_{i=0}^2 c_i \sum_\alpha
\left\{\sqrt{\epsilon_\alpha^2+\Lambda_i^2}
-\sqrt{\epsilon_\alpha^{(0)2}+\Lambda_i^2}
\right\}
+m_\pi^2f_\pi^2\int d^3r \, \left[1-{\rm cos}(\Theta)\right]\, .
\label{etot}
\end{equation}
Here we have also subtracted the vacuum energy associated with the non-dynamical 
meson field configuration $\Theta\equiv0$ (denoted by the superscript) that is 
often called the cosmological constant contribution. This subtraction will also 
play an important role for the unpolarized isosinglet structure function as it enters
via the momentum sum rule. The soliton profile is then obtained as the profile 
function $\Theta(r)$ that minimizes the total energy $E_{\rm tot}$ self-consistently 
subject to the above mentioned boundary conditions.

This soliton represents an object which has unit baryon number but neither good quantum 
numbers for spin and flavor (isospin). Such quantum numbers are generated by canonically 
quantizing the time-dependent collective coordinates $A(t)$ which parameterize the 
spin-flavor orientation of the soliton. For a rigidly rotating soliton the Dirac operator 
becomes, after transforming to the flavor rotating frame \cite{Reinhardt:1989st},
\begin{equation}
\imu\mathbf{D}^{(\pi)}=A\beta\left(\imu\partial_t - 
\Vek{\Omega}\cdot\Vek{\tau}-h\right)A^\dagger
\quad{\rm and}\quad
\imu\mathbf{D}^{(\pi)}_5=A\left(-\imu\partial_t + 
\Vek{\Omega}\cdot\Vek{\tau} -h\right)\beta A^\dagger\, .
\label{collq1}
\end{equation}
Actual computations involve an expansion with respect to
the angular velocities 
\begin{equation}
A^\dagger \frac{d}{dt} A = \frac{\imu}{2}\Vek{\Omega}\cdot\Vek{\tau}\,,
\label{collq2}
\end{equation}
which, according to the quantization rules, are replaced by the spin operator 
\begin{equation}
\Vek{\Omega}\longrightarrow \frac{1}{\alpha^2}\, \Vek{J}\, .
\label{collq3}
\end{equation}
The constant of proportionality is the moment of inertia 
\begin{equation}
\alpha^2=\frac{N_C}{4}\left[1+{\rm sign}(\epsilon_{\rm v})\right]
\sum_{\beta\ne{\rm v}} \frac{|\langle{\rm v}|\tau_3|\beta\rangle|^2}
{\epsilon_{\beta}-\epsilon_{\rm v}}
+\frac{N_C}{8}\sum_{\alpha\ne\beta}\sum_{i=1}^2c_i
\frac{|\langle\alpha|\tau_3|\beta\rangle|^2}{\epsilon_\alpha^2-\epsilon_\beta^2}
\left\{\frac{\epsilon_\alpha^2+\epsilon_\alpha\epsilon_\beta+2\Lambda_i^2}
{\sqrt{\epsilon_\alpha^2+\Lambda_i^2}}
-\frac{\epsilon_\beta^2+\epsilon_\alpha\epsilon_\beta+2\Lambda_i^2}
{\sqrt{\epsilon_\beta^2+\Lambda_i^2}}\right\}\,,
\label{mominert}\end{equation}
which is of the order $N_C$. With Eq.~(\ref{collq3}) 
the expansion in $\Vek{\Omega}$ is thus equivalent 
to the one in $1/N_C$. The nucleon wave-function becomes a (Wigner D) 
function of the collective coordinates. A useful relation in 
computing matrix elements of nucleon states is \cite{Adkins:1983ya}
\begin{equation}
\langle N |\frac{1}{2}{\rm tr}
\left(A^\dagger\tau_i A\tau_j\right) |N\rangle =
-\frac{4}{3}\langle N | I_i J_j | N\rangle\, .
\label{collq4}
\end{equation}

\section{Structure functions from soliton}
\label{sol_st_fct}

We first repeat the relation between the structure functions and the 
hadronic tensor of the nucleon. Its symmetric combination
$W_{\mu\nu}^S(q)=\frac{1}{2} \left(W_{\mu\nu}(q) + W_{\nu\mu}(q)\right)$,
is parameterized by two form factors\footnote{Factors of the nucleon mass, $M$,
occur for dimensional reasons.}
\begin{align}
W_{\mu \nu}^S(q) & = M W_1(\nu,Q^{2}) \left( -g_{\mu\nu}
+ \frac{q_\mu q_\nu}{q^2} \right)+ \frac{W_2(\nu,Q^2)}{M}
\left( p_\mu-\frac{p\cdot q}{q^2} q_\mu \right)
\left( p_\nu - \frac{p \cdot q}{q^2} q_\nu \right)\,.
\end{align}
In the Bjorken limit, Eq.~(\ref{eq:defBlim}), these form factors turn into the 
unpolarized structure functions that we extract by appropriate projections:
\begin{equation}
F_1(x) = -\frac{1}{2}g^{\mu\nu} W_{\mu\nu}^S(q)
\qquad {\rm and}\qquad  
\quad F_2(x)=-xg^{\mu\nu}W_{\mu\nu}^S(q).
\end{equation}
It must be noted that the Callan Gross relation, i.e. $F_1(x)=2xF_2(x)$, is satisfied
in this case by construction. Similarly, the anti-symmetric part is also parameterized by 
two form factors
\begin{align}
W_{\mu \nu}^A(q) & = \imu \epsilon_{ \mu \nu \lambda \sigma} q^{\lambda}
\left\lbrace M G_{1} (\nu, Q^{2}) s^{\sigma} + \frac{G_{2}(\nu, Q^{2})}{M}
\left( (p \cdot q) s^{\sigma} - (q \cdot s) p^{\sigma} \right)  \right\rbrace.
\end{align}
In the Bjorken limit these form factors yield the structure functions
\begin{align}
\mathsf{g}_1(x)= M^{2}\nu G_1(\nu, Q^2)
\qquad {\rm and}\qquad \mathsf{g}_2(x) = M \nu G_2(\nu, Q^2)\,.
\end{align}
The longitudinal, $\mathsf{g}_{1}(x)$, and transverse,
$\mathsf{g}_{T}(x)=\mathsf{g}_{1}(x) + \mathsf{g}_{2}(x)$, structure functions are 
extracted from the hadronic tensor using the projection operators
\begin{align}
\mathsf{g}_{1}(x) = \frac{\imu}{2M} \epsilon^{ \mu \nu \rho \sigma}
\frac{q_{\sigma} p_{\rho}}{q \cdot s} W_{\mu\nu}^{A}(q) \quad \text{and} \quad \Vek{q}
\parallel \Vek{s} \\
\mathsf{g}_{T}(x) = \frac{-\imu }{2 M} \epsilon^{ \mu \nu \rho \sigma} s_{\rho} p_{\sigma}
W_{\mu\nu}^{A}(q) \quad \text{and} \quad \Vek{q} \perp \Vek{s}.
\end{align}

To obtain the hadronic tensor for the nucleon in the soliton model, the functional traces in 
Eqs.~(\ref{simple6}) and~(\ref{simple7}) are computed using the basis defined by the 
self-consistent soliton, Eq.~(\ref{diagh}). This calculation has been detailed in 
Ref.~\cite{Weigel:1999pc} that we adopt directly. We start with the leading 
order in $\frac{1}{N_C}$ to the vacuum (or sea) contribution to~$W_{\mu\nu}$
\begin{align}
W^{\rm (s)}_{\mu\nu}(q) & = -\imu \frac{ M N_c\pi}{8} \int \frac{ d\omega}{2\pi}  
\sum_\alpha \int d^3 \xi \int \frac{d\lambda}{2 \pi}\, 
{\rm e}^{\imu Mx \lambda} 
\label{lead_had_tensor}\\
&\hspace{0.8cm}
\times\Big\langle N,\Vek{s} \Big| \Bigl\{ \Bigl[ 
\overline{\Psi}_\alpha(\Vek{\xi}) \mathcal{Q}_A^2 \gamma_\mu \nslash 
\gamma_\nu \Psi_\alpha (\Vek{\xi} + \lambda \hat{e}_3) 
{\rm e}^{-\imu \lambda \omega}
-\overline{\Psi}_\alpha (\Vek{\xi}) \mathcal{Q}_A^2 \gamma_\nu \nslash
\gamma_\mu \Psi_\alpha (\Vek{\xi} - \lambda\hat{e}_3) {\rm e}^{\imu \lambda \omega} 
\Bigr] \left. f_\alpha^+ (\omega)\right|_{\rm p} \cr
&\hspace{1.2cm}+\Bigl[\overline{\Psi}_\alpha (\Vek{\xi}) \mathcal{Q}_A^2 
\left(\gamma_\mu \nslash \gamma_\nu \right)_5 \Psi_\alpha 
(\Vek{\xi} - \lambda \hat{e}_3) {\rm e}^{-\imu \lambda \omega} 
- \overline{\Psi}_\alpha (\Vek{\xi} ) \mathcal{Q}_A^2  
\left( \gamma_\nu \nslash \gamma_\mu \right)_5 
\Psi_\alpha (\Vek{\xi} + \lambda \hat{e}_3) {\rm e}^{\imu \lambda \omega}\Bigr] 
\left. f_\alpha^- (\omega)\right|_{\rm p}
\Bigr\} \Big| N,\Vek{s} \hspace{0.1cm} \Big\rangle\,.
\nonumber\end{align} 
Here $n^\mu=\left(1,0,0,1\right)^{\mu}$ is the light-cone vector of the 
photon momentum while $\mathcal{Q}_A=A^\dagger \mathcal{Q} A$ denotes
the flavor rotated quark charge matrix from which we compute nucleon 
matrix elements as in Eq.~(\ref{collq4}). Furthermore
\begin{equation}
f_\alpha^\pm = \sum_{i=0}^{2} c_i \frac{\omega \pm \epsilon_\alpha}
{\omega^2 - \epsilon_\alpha^2 - \Lambda_i^2 + \imu \epsilon} \pm 
\frac{\omega \pm \epsilon_\alpha}{\omega^2 - \epsilon_\alpha^2 +\imu \epsilon}, 
\label{spect_func}
\end{equation} 
are Pauli-Villars regularized spectral functions. The subscript `$p$' indicates 
their pole contributions that we will explain below.

For the vacuum contribution to the isosinglet unpolarized structure function we then 
obtain
\begin{align}
\left[F_1^{I=0}(x)\right]_s & = \imu \frac{5 \pi}{72} M N_c  \int \frac{d\omega}{2 \pi} 
\sum_\alpha \int \frac{d\lambda}{2 \pi} {\rm e}^{\imu Mx \lambda} \left( \sum_{i=0}^2 c_i 
\frac{\omega + \epsilon_\alpha}{\omega^2 - \epsilon_\alpha^2 - \Lambda_i^2 
+ \imu  \epsilon} \right)_{\rm p} \cr
& \times \int d^3 \xi \, \Bigl\{ \Psi_\alpha^\dagger (\Vek{\xi}) 
(1 - \alpha_3) \Psi_\alpha (\Vek{\xi} + \lambda \hat{e}_3) {\rm e}^{-\imu \omega \lambda} 
- \Psi_\alpha^\dagger (\Vek{\xi}) (1 -\alpha_3) \Psi_\alpha 
(\Vek{\xi} - \lambda \hat{e}_{3}) {\rm e}^{\imu \omega \lambda}  \Bigr\}. 
\label{unpolar_function}
\end{align} 
Here the pole contributions is
\begin{align}
\left( \sum_{i=0}^2c_i \frac{1}{\omega^2-\epsilon_\alpha^2
-\Lambda_i^2+\imu\epsilon} \right)_{\rm p} & =  \sum_{i=0}^2 c_i
\frac{-\imu \pi}{ \omega_\alpha } \left[ \delta \left( \omega + \omega_\alpha \right) 
+ \delta \left( \omega -\omega_\alpha \right) \right]\,,
\label{eq:pol1} \end{align} 
where (eventually we take the single cut-off limit as described after 
Eq.~(\ref{pvcond}) and thus omit the label $i$ on $\omega_\alpha$)
\begin{equation}
\omega_\alpha = \sqrt{\epsilon_\alpha^2 + \Lambda_i^2}\,.
\end{equation} 
We recall that the single cut-off approach requires a derivative with 
respect to that cut-off. Of course, this also affects the implicit dependence 
of $\omega_\alpha$ on that cut-off.

We introduce the Fourier transform of the quark wave-function as
\begin{equation}
\widetilde{\Psi}_\alpha (\Vek{p}) = \int \frac{d^3 r}{4 \pi}\, \Psi_\alpha(\Vek{r}) 
{\rm e}^{\imu \Vek{r} \cdot \Vek{p}}. \label{Fourier_trans}
\end{equation} 
Implementing a full Fourier transform differs from the approaches of Refs. 
\cite{Wakamatsu:1997en,Pobylitsa:1998tk} who used the expansion from diagonalizing 
the Dirac Hamiltonian, Eq.~\eqref{diagh}. This resulted in discontinuities of the 
numerically computed quark distributions and required a smoothening procedure.

Performing the frequency ($\omega$) and lambda\footnote{Technically it is 
advisable to average the photon direction rather than fixing it along 
the $z$-axis \cite{Diakonov:1997vc}.} `$\lambda$' integrals gives the 
vacuum contribution of the flavor-singlet unpolarized structure function in the 
nucleon rest frame (RF)
\begin{align}
\left[ F_1^{I=0} (x)\right]_{\rm s}^{\mp} & = \frac{5MN_{c}}{144} \sum_\alpha 
\sum_{i=0}^{2} c_i  \int_{\vert M  x_\alpha^{\pm} \vert}^{\infty} p\, d p\,  
\int d \Omega_p \Biggl\{ \pm  \widetilde{\Psi}_\alpha^\dagger (\Vek{p}) 
\widetilde{\Psi}_\alpha (\Vek{p}) - \frac{\epsilon_\alpha}{ \omega_\alpha} 
\frac {M x_\alpha^{\pm}}{p} \widetilde{\Psi}_\alpha^{\dagger} (\Vek{p}) 
\hat{\Vek{p}} \cdot \Vek{\alpha} \widetilde{\Psi}_\alpha (\Vek{p}) \Biggr\}, 
\label{vacuum_cont_F1}
\end{align} 
where   
\begin{equation}
M x_\alpha^{\pm}  = M x \pm \omega_\alpha. \label{def_of_Mx}
\end{equation} 
In the above $[F_1^{I=0}(x)]_{\rm s}^{\pm}$ refers to the positive (negative) 
frequency components that are typically referred to as quark and antiquark distributions. 
In our calculation they arise from the two poles of the $\delta$-function in 
Eq.~(\ref{eq:pol1}). Then the vacuum part of the isoscalar, unpolarized structure function 
becomes
\begin{equation}
[F_1^{I=0}(x)]_{\rm s} = [F_1^{I=0}(x)]_{\rm s}^{-} + [F_1^{I=0}(x)]_{\rm s}^{+}.
\label{eq:f1s}
\end{equation}
As a matter of fact, this is still not the full result. Substituting free spionors 
(not interaction with the soliton) produces a non-zero result.  This non-zero result 
must also be subtracted. In the discussion of the sum rules below we will see that 
this is nothing but the $\epsilon^{(0)}_\alpha$ type subtraction performed in 
Eq.~(\ref{etot}) and may be considered a cosmological constant type contribution.

The valence quark contribution is obtained by replacing the quark levels in 
\eqref{vacuum_cont_F1} by the cranked valence level 
\begin{align}
\Psi_{\rm v}^{(\rm rot)} (\Vek{r},t) & = {\rm e}^{-\imu \epsilon_{\rm v}t} A(t) 
\Biggl\{ \Psi_{\rm v}(\Vek{r}) + \frac{1}{2} \sum_{\alpha \neq \rm v} 
\Psi_\alpha (\Vek{r}) \frac{ \langle \alpha | \Vek{\tau} \cdot \Vek{\Omega} 
| \rm v \rangle }{ \epsilon_{\rm v}-\epsilon_\alpha}\Biggr\} 
={\rm e}^{-\imu \epsilon_{\rm v}t} A(t) \psi_{\rm v}(\Vek{r})\,,
\label{cranked_valence_level}
\end{align}
In the above $\psi_{\rm v} (\Vek{r})$ is the spatial part of the valence quark 
wave-function with the rotational correction included and $\epsilon_{\rm v}$ is 
the energy eigenvalue of the valence quark level. Noting that the valence quark 
wave-function has positive parity and the pole contribution $ \left. f^{\pm} 
\right|_{\rm pole} = -4 \imu \pi \delta \left( \omega \mp \epsilon_{\rm v}\right) $ 
gives the valence quark contribution 
\begin{align}
\left[ F_1^{I=0}(x) \right]_{\rm v}^{\mp}  & = -\frac{5 M N_c}{144}   
\left[1+{\rm sign}(\epsilon_{\rm v})\right]
 \int_{M \vert x_{\rm v}^{\pm} \vert}^{\infty} p\, d p \int d \Omega_p 
\Biggl\{ \pm  \widetilde{\Psi}_{\rm v}^{\dagger} (\Vek{p}) \widetilde{\Psi}_{\rm v} 
(\Vek{p}) - \frac{M x_{\rm v}^{\pm}}{p}  \widetilde{\Psi}_{\rm v}^{\dagger} 
(\Vek{p}) \hat{\Vek{p}} \cdot \Vek{\alpha} \widetilde{\Psi}_{\rm v} (\Vek{p})\Biggr\}, 
\label{valence_cont_F1}
\end{align}
where
$M x_{\rm v}^{\pm} = M x \pm \epsilon_{\rm v}$.
Again, we have separated positive and negative frequency components.

The quark spinors $\Psi_\nu(\Vek{r})$ separate into radial and angular pieces \cite{Kahana:1984be}. 
At the end, the structure functions, as in Eq.~(\ref{vacuum_cont_F1}) are computed as integrals 
over the (Bessel-)Fourier transforms of the radial functions in the quark spinors. In 
Appendix \ref{app:sol} we list examples explicitly.

In quite an analogous manner, the isovector components of the polarized structure functions 
are extracted from the anti-symmetric combination $W_{\mu\nu}^A(q)$.  Explicitly we find
the vacuum contribution to the longitudinal polarized structure function to be
\begin{align}
\left[\mathsf{g}_{1}^{I=1}(x)\right]_{\rm s}^{\mp}&= 
- \frac{M N_{c}}{72} I_{3} \sum_{\alpha} \sum_{i=0}^{2} c_{i}  
\Biggl\{ \mp \int_{\vert M x_{\alpha}^{\pm } \vert}^{\infty} 
 d p\,M x_{\alpha}^{\pm} \int d \Omega_{p}   
\widetilde{\Psi}_{\alpha}^{\dagger} (\Vek{p}) \hat{\Vek{p}} \cdot \Vek{\tau} 
\gamma_{5} \widetilde{\Psi}_{\alpha} (\Vek{p}) \cr
&  - \frac{\epsilon_{\alpha}}{\omega_\alpha} 
\int_{\vert M x_{\alpha}^{\pm} \vert}^{\infty} d p\,p^{2} 
\Biggl[ A_{\pm} \int d \Omega_{p}  
\widetilde{\Psi}_{\alpha}^{\dagger} (\Vek{p}) \Vek{\tau} 
\cdot \Vek{\sigma} \widetilde{\Psi}_{\alpha} (\Vek{p})   
+  B_{\pm}  \int d \Omega_{p} \widetilde{\Psi}_{\alpha}^{\dagger} (\Vek{p}) 
\hat{\Vek{p}} \cdot \Vek{\tau} \hat{\Vek{p}} \cdot \Vek{\sigma}\widetilde{\Psi}_{\alpha} 
(\Vek{p}) \Biggr] \Biggr\}, \label{vac_cont_isovec_longpolar}
\end{align}
and the isovector transverse polarized structure function as 
\begin{align}
\left[\mathsf{g}_{T}^{I=1}(x)\right]_{\rm s}^{\mp} &=\frac{M N_{c}}{144} I_{3}
\sum_{\alpha} \sum_{i=0}^{2} c_{i} \Biggl\{ \frac{\epsilon_{\alpha}}{\omega_\alpha} 
\int_{\vert M x_{\alpha}^{\pm} \vert}^{\infty} d p\,p  \int d \Omega_{p}  
\widetilde{\Psi}_{\alpha}^{\dagger} (\Vek{p}) \Vek{\tau} \cdot \Vek{\sigma} 
\widetilde{\Psi}_{\alpha} (\Vek{p}) \cr
& -\frac{\epsilon_{\alpha}}{\omega_\alpha} \int_{\vert M x_{\alpha}^{\pm} \vert}^{\infty} 
d p\,p^{2} \Biggl[A_{\pm} \int d \Omega_{p}  \widetilde{\Psi}_{\alpha}^{\dagger} (\Vek{p}) 
\Vek{\tau} \cdot \Vek{\sigma} \widetilde{\Psi}_{\alpha} (\Vek{p})+ B_{\pm} \int d \Omega_{p} 
\widetilde{\Psi}_{\alpha}^{\dagger} (\Vek{p}) \hat{\Vek{p}} \cdot \Vek{\tau} \hat{\Vek{p}} \cdot 
\Vek{\sigma}\widetilde{\Psi}_{\alpha} (\Vek{p})\Biggr]\Biggr\}.
\label{vac_cont_isovec_transpolar}
\end{align}
In these formulas we have introduced the abbreviations, see also Eq.~\eqref{def_of_Mx}
\begin{align}
A_{\pm} & = \frac{1}{2p} \left( 1 - \frac{(Mx_{\alpha}^{\pm})^{2}}{p^{2}}\right),
\quad  B_{\pm}  = \frac{1}{2p} \left( 3\frac{(Mx_{\alpha}^{\pm})^{2}}{p^{2}} -1 \right).
\label{long_multiple_factors}
\end{align}
The total (vacuum contribution to the) polarized structure functions is the sum of the 
positive ($+$) and negative ($-$) frequency components. Again, some details in terms of 
the Fourier transformed radial functions are given in Appendix~\ref{app:polSF}. For 
completeness we also list the formulas for the valence quark contribution \cite{Weigel:1996jh}. 
The contribution to the longitudinal polarized structure function is obtained as
\begin{align}
\left[\mathsf{g}_{1}^{I=1}(x)\right]_{\rm v}^{\mp}& = \frac{M N_{c}}{72} 
\left[1+{\rm sign}(\epsilon_{\rm v})\right]I_{3}  
\Biggl\{\mp \int_{\vert M x^{\pm} \vert}^{\infty} d p\,M x_{\rm v}^{\pm} 
\int d \Omega_{p} \widetilde{\Psi}_{\rm v}^{\dagger} (\Vek{p})\hat{\Vek{p}}\cdot \Vek{\tau} 
\gamma_{5} \widetilde{\Psi}_{\rm v} (\Vek{p}) \cr 
& -\int_{\vert M x_{\rm v}^{\pm} \vert}^{\infty} d p\,p^{2} 
\Biggl[ A_{\pm} \int d \Omega_{p} \widetilde{\Psi}_{\rm v}^{\dagger} 
(\Vek{p}) \Vek{\tau} \cdot \Vek{\sigma} \widetilde{\Psi}_{\rm v} (\Vek{p})  
+ B_{\pm} \int d \Omega_{p} \widetilde{\Psi}_{\rm v}^{\dagger} (\Vek{p}) \hat{\Vek{p}} \cdot 
\Vek{\tau} \hat{\Vek{p}} \cdot \Vek{\sigma} \widetilde{\Psi}_{\rm v} (\Vek{p}) \Biggr] \Biggr\},
\label{val_cont_isovec_longpolar}
\end{align}
and that for the transverse polarized structure function as 
\begin{align}
\left[\mathsf{g}_{T}^{I=1}(x)\right]_{\rm v}^{\mp} & = \frac{M N_{c}}{144}
\left[1+{\rm sign}(\epsilon_{\rm v})\right] I_{3} 
\Biggl\{ \int_{\vert M x_{\rm v}^{\pm} \vert}^{\infty} d p\,p  \int d \Omega_{p}   
\widetilde{\Psi}_{\rm v}^{\dagger} (\Vek{p}) \Vek{\tau} \cdot \Vek{\sigma} 
\widetilde{\Psi}_{\rm v} (\Vek{p})  \cr
& -\int_{\vert M x_{\rm v}^{\pm} \vert}^{\infty} d p\,p^{2} 
\Biggl[ A_{\pm}\int d \Omega_{p} \widetilde{\Psi}_{\rm v}^{\dagger} 
(\Vek{p}) \Vek{\tau} \cdot \Vek{\sigma} \widetilde{\Psi}_{\rm v} (\Vek{p})  
+ B_{\pm}\int d \Omega_{p} \widetilde{\Psi}_{\rm v}^{\dagger} (\Vek{p}) 
\hat{\Vek{p}} \cdot \Vek{\tau} \hat{\Vek{p}} \cdot \Vek{\sigma}   
\widetilde{\Psi}_{\rm v} (\Vek{p}) \Biggr] \Biggr\}.
\label{val_cont_isovec_transpolar}
\end{align}
The isovector unpolarized and isoscalar polarized structure functions are subleading 
in the $1/N_C$ counting. They are also more complicated to compute as they are quartic
in the spinors and involve double sums over the basis states defined Eq.~(\ref{diagh}).
We refrain from presenting those lengthy expressions here and rather refer the interested 
reader to the Appendixes of Ref.~\cite{TakyiThesis}.

\section{Formal discussion of sum rules}
\label{sec:SR}

In this section we discuss how the sum rules for the unpolarized and polarized structure 
functions work out when written explicitly in terms of the momentum space eigenspinors 
$\widetilde{\Psi}_\alpha$. In this context it is important to note that we compute 
the structure functions for a localized configuration in its rest frame. Then the 
Bjorken variable has support on the half axis from zero to infinity. Lorentz covariance
is regained by transforming to the infinite momentum frame, {\it cf.} Section \ref{ssub:pande}.

Sum rules relate integrated structure functions to static observables. In soliton models 
the latter are directly expressed in terms of the eigenspinors, Eq.~(\ref{diagh}) in 
coordinates space. Typically the sum rules can then be expressed as level-by-level identities. 
The only exception is the momentum (or energy) sum rule. For it to be obeyed it is compulsory 
that the soliton is an extremum of the energy functional, Eq.~(\ref{etot}).

\subsection{Momentum sum rule}
For the momentum sum rule we require that $\frac{36}{5}\int dx x F_1(x)$ produces
the quark contribution to the classical energy, {\it i.e.} all but the last integral 
in Eq.~(\ref{etot}). First we consider the scalar terms, 
$\pm\widetilde{\Psi}_{\alpha}^{\dagger} (\Vek{p}) \widetilde{\Psi}_{\alpha}(\Vek{p})$,
from the vacuum contribution, Eq.~(\ref{vacuum_cont_F1})\footnote{We adopt the 
notation
$ \displaystyle 
\Big\langle \alpha \Big|\hat{O}\Big|\alpha \Big\rangle_{a} 
= \int_{a}^{\infty} d p\,p \int d \Omega_p 
\widetilde{\Psi}^{\dagger}_{\alpha}(\Vek{p})\hat{O}\widetilde{\Psi}_{\alpha} (\Vek{p})$.} 
\begin{align}
\left[\mathcal{M}_{G}^{0} \right]_{\rm s} & 
= \frac{M N_c}{4} \sum_{\alpha} \sum_{i=0}^{2} c_i \int_{0}^{\infty} d x\,x 
\Biggl\{\Big\langle \alpha \Big|\alpha\Big\rangle_{\vert M x^+\vert} 
-\Big\langle \alpha \Big|\alpha\Big\rangle_{\vert M x^-\vert }\Biggr\}\cr 
&=\frac{M N_c}{4}\sum_\alpha \sum_{i=0}^{2} c_i 
\Biggl\{\int_{\frac{\omega_\alpha}{M }}^{\infty}d y\,\left(y -\frac{\omega_\alpha}{M }\right) 
\Big\langle\alpha \Big|\alpha\Big\rangle_{ M y } 
-\int_{-\frac{\omega_\alpha}{M }}^{\infty} dy\,\left(y + \frac{\omega_\alpha}{M }\right) 
\Big\langle \alpha \Big| \alpha \Big\rangle_{\vert M y\vert} \Biggr\}, \cr 
& =-\frac{N_c}{2 }\sum_{\alpha} \sum_{i=0}^{2} c_i\, \omega_\alpha \int_{0}^{\infty} d y\, 
\Big\langle \alpha \Big| \alpha \Big\rangle_{ M y }
=  \frac{N_c}{2 } 
\sum_\alpha \sum_{i=0}^{2} c_i\,\omega_\alpha \int_{0}^{\infty} d y\, 
y\frac{\partial}{\partial y} \int_{My}^{\infty} d p\,p  \int d \Omega 
\widetilde{\Psi}^{\dagger}_{\alpha}(\Vek{p})\widetilde{\Psi}_\alpha(\Vek{p}), \cr
& =-\frac{M N_c}{2} \sum_\alpha \sum_{i=0}^{2} c_i\, \omega_\alpha \int_{0}^{\infty}
d y\, y \left[ p \int d\Omega_p 
\widetilde{\Psi}_{\alpha}^{\dagger}(\Vek{p})
\widetilde{\Psi}_\alpha(\Vek{p})\right]_{p=My} =-\frac{N_c}{2M}\sum_\alpha\sum_{i=0}^{2} 
c_i \sqrt{\epsilon_{\alpha}^{2}+\Lambda_{i}^{2}},
\end{align}
which is $1/M$ times the vacuum contribution to the classical 
energy. This contribution also includes subtraction of the trivial vacuum 
energy, when there is no soliton. Hence the isoscalar unpolarized structure 
function necessitates the analog subtraction, as indicated earlier.
For the valence contribution the momentum sum rule gives
\begin{equation}
\left[\mathcal{M}_{G}^{0} \right]_{\rm v} = \frac{36}{5} \int_{0}^{\infty} dx\, 
x\left[F_{1}^{I=0}(x)\right]_{\rm v} 
=\frac{N_C}{2M}\left[1+{\rm sign}(\epsilon_{\rm v})\right] \epsilon_{\rm v}\,.
\end{equation}
Similarly, integrating the term with the operator
$\hat{\Vek{p}} \cdot \Vek{\alpha} $ gives 
\begin{equation}
\left[ \mathcal{M}_{G}^{1} \right]_{\rm s} 
=-\frac{M^2 N_c}{2}\sum_\alpha \sum_{i=0}^{2} c_i \frac{\epsilon_\alpha}{\omega_\alpha} 
\int_{0}^{\infty} d y\,y^2 \Big\langle\alpha\Big|\frac{1}{p}\hat{\Vek{p}}\cdot\Vek{\alpha}\Big|\alpha 
\Big\rangle_{ M y}
=-\frac{N_c}{6M}\sum_\alpha\sum_{i=0}^{2}c_i\frac{\epsilon_\alpha}{\omega_\alpha}
\langle\alpha|\Vek{\alpha}\cdot{\Vek{p}}|\alpha\rangle\,.
\end{equation} 
Next we use the Dirac Hamiltonian, Eq.~(\ref{hedgehog}) to write
$$
\imu\Vek{\alpha}\cdot\Vek{p}=\left[\Vek{r}\cdot\Vek{p},h\right]
-m\beta\left[\Vek{r}\cdot\Vek{p},U_5(\Vek{r})\right]
$$
so that 
$\langle\alpha|\Vek{\alpha}\cdot{\Vek{p}}|\alpha\rangle=
\imu m\beta \langle\alpha|\left[\Vek{r}\cdot\Vek{p},U_5(\Vek{r})\right]|\alpha\rangle$.
Since $\Vek{r}\cdot\Vek{p}$ is the dilatation operator this matrix element 
measures the change of the single particle energy when scaling the soliton extension 
by an infinitesimal amount. Furthermore $\frac{\epsilon_\alpha}{\omega_\alpha}=
\frac{\partial}{\partial\epsilon_\alpha} \sqrt{\epsilon^2_\alpha+\Lambda^2_i}$ so that 
$\left[\mathcal{M}_{G}^{1}\right]_{\rm s}$ is the change of the vacuum energy 
when the soliton extension deviates slightly from its stationary point. 
Similarly, the valence quark adds
$\displaystyle \left[\mathcal{M}_{G}^{1}\right]_{\rm v}=\imu\frac{N_C}{12M} 
\left[1+{\rm sign}(\epsilon_{\rm v})\right]
\left\langle {\rm v}|\left[\Vek{r}\cdot\Vek{p},U_5(\Vek{r})\right]|{\rm v}\right\rangle$
to the sum rule. Then
$\left[\mathcal{M}_{G}^{1}\right]_{\rm s} 
+ \left[\mathcal{M}_{G}^{1} \right]_{\rm v}$ is the coefficient of $(\lambda-1)$ 
term in the expansion 
\begin{align}
E[U(\lambda \Vek{x})]=E_0 +(\lambda-1)E_1 +\cdots(\lambda-1)^l E_l + \cdots
\end{align}
of the classical energy. Since $U(\Vek{x})$ is a stationary point, $E_{1}=0$ 
thus verifying the momentum sum rule \cite{Diakonov:1996sr}. Obviously the 
momentum sum rule is not saturated level by level; rather it requires summing 
all contributions to this isoscalar unpolarized structure function. Hence this 
sum rule will be a very sensitive test of the numerical simulation.

\subsection{Bjorken sum rule}
Here we verify the Bjorken sum rule in our model, which relates the 
isovector polarized structure function $\mathsf{g}_{1}^{I=1}$
to the axial charge \cite{Bjorken:1969mm}. First, we show that the term 
in Eq.~(\ref{vac_cont_isovec_longpolar}) with the operator 
$\hat{\Vek{p}}\cdot\Vek{\tau}\gamma_5$ integrates to zero
\begin{align}
&\sum_\alpha \sum_{i=0}^{2} c_i \int_{0}^{\infty}d x\, 
\Biggl\{M x^- \Big\langle\alpha\Big|\frac{1}{p}\hat{\Vek{p}}\cdot\Vek{\tau}\gamma_5 \Big|\alpha 
\Big\rangle_{\vert M x^- \vert}-M x^+ \Big\langle\alpha\Big|\frac{1}{p}\hat{\Vek{p}}\cdot
\Vek{\tau}\gamma_5 \Big|\alpha\Big\rangle_{\vert M x^+ \vert}\Biggr\}, \cr
&\hspace{1cm}=\sum_\alpha \sum_{i=0}^{2} c_{i}  \Biggl\{ 
\int_{-\frac{\omega_0}{M}}^{\infty}d y\, My \Big\langle\alpha\Big|\frac{1}{p}\hat{\Vek{p}}\cdot 
\Vek{\tau}\gamma_5 \Big|\alpha\Big\rangle_{\vert My\vert} 
-\int_{\frac{\omega_0}{M }}^{\infty}d y\, My\Big\langle\alpha 
\Big|\frac{1}{p}\hat{\Vek{p}}\cdot\Vek{\tau} \gamma_5 \Big|\alpha\Big\rangle_{My}\Biggr\},\cr
&\hspace{1cm}=\sum_\alpha \sum_{i=0}^{2} c_i 
\int_{-\frac{\omega_0}{M}}^{\frac{\omega_0}{M }} d y\, My\Big\langle 
\alpha\Big|\frac{1}{p}\hat{\Vek{p}}\cdot\Vek{\tau} \gamma_5 \Big|\alpha\Big
\rangle_{\vert M y \vert}=0\,. 
\label{bjor_sum_lead_order} \end{align} 
There are two contributions without $\gamma_{5}$. The first one contributes
\begin{align}
&\frac{M N_c I_3}{72}\sum_\alpha\sum_{i=0}^{2} c_i 
\frac{\epsilon_\alpha}{\omega_\alpha}\int_{0}^{\infty} d x\,\Biggl[
\Big\langle\alpha \Big|pA_{+}\Vek{\tau}\cdot\Vek{\sigma}\Big|\alpha
\Big\rangle_{\vert M x^+ \vert }
+\Big\langle \alpha\Big|pA_{-}\Vek{\tau}\cdot\Vek{\sigma}\Big|\alpha
\Big\rangle_{\vert M x^- \vert}\Biggr]\cr
&\hspace{1cm}
=\frac{M N_c I_3}{144}\sum_\alpha \sum_{i=0}^{2}c_i \frac{\epsilon_\alpha}{\omega_\alpha} 
\Biggl[\int_{\frac{\omega_0}{M}}^{\infty}d y\Big\langle\alpha\Big|\Vek{\tau} 
\cdot\Vek{\sigma}\left(1-\frac{(My)^{2}}{p^{2}}\right)\Big|\alpha\Big\rangle_{ M y}
+\int_{-\frac{\omega_0}{M }}^{\infty} dy \Big\langle\alpha\Big|
\Vek{\tau}\cdot\Vek{\sigma}\left(1-\frac{(My)^{2}}{p^2}\right)
\Big|\alpha\Big\rangle_{\vert M y \vert }\Biggr]\cr
&\hspace{1cm}=\frac{N_c I_3}{108} \sum_{i=0}^{2} c_i \frac{\epsilon_\alpha}{\omega_\alpha}\Big\langle 
\alpha|\Vek{\tau}\cdot\Vek{\sigma}|\alpha\Big\rangle\,. 
\label{bjor_sum_sub_lead_order} \end{align}
The term with $\hat{\Vek{p}} \cdot\Vek{\tau} \hat{\Vek{p}} \cdot \Vek{\sigma}$ disappears because
\begin{align}
&\int_{0}^{\infty} d y\int_{My}^{\infty} p^{2} d p \int d\Omega_p 
\left(\frac{1}{p}-3\frac{(My)^2}{p^3}\right)
\widetilde{\Psi}_{\alpha}^{\dagger}(\Vek{p})\hat{\Vek{p}}\cdot\Vek{\tau}\hat{\Vek{p}}\cdot\Vek{\sigma} 
\widetilde{\Psi}_\alpha (\Vek{p})\cr
&\hspace{3cm}=\int_{0}^{\infty} dy\int_{My}^{\infty} dp \int d\Omega_p
\frac{\partial}{\partial y}\left(py-\frac{M^2y^3}{p}\right)
\widetilde{\Psi}_{\alpha}^{\dagger}(\Vek{p})\hat{\Vek{p}}\cdot\Vek{\tau}\hat{\Vek{p}}\cdot\Vek{\sigma}
\widetilde{\Psi}_\alpha (\Vek{p})\cr
&\hspace{3cm}=\int_{0}^{\infty} d y \left(\frac{M^2y^3}{p}-py\right) 
\left[\int d\Omega_p \widetilde{\Psi}_{\alpha}^{\dagger}(\Vek{p})\hat{\Vek{p}}\cdot\Vek{\tau} 
\hat{\Vek{p}}\cdot\Vek{\sigma}\widetilde{\Psi}_{\alpha}(\Vek{p})\right]_{p=My}=0\,.
\hspace{2cm}\label{B_cont} \end{align}
Hence the Bjorken sum rule for the vacuum contribution of the longitudinal polarized 
structure function becomes
\begin{equation}
\int d x \left[\mathsf{g}_{1}^{p}(x) -\mathsf{g}_{1}^{n}(x)\right]_{\rm s} 
=\frac{N_c }{108} \sum_{i=0}^{2} c_i \frac{\epsilon_\alpha}{\omega_0} 
\Big\langle\alpha\Big|\Vek{\tau}\cdot\Vek{\sigma}\Big|\alpha \Big\rangle 
=\frac{1}{6}\left[\frac{N_C }{18} \sum_{\alpha} \sum_{i=0}^{2} c_{i} 
\frac{\epsilon_\alpha}{\sqrt{\epsilon_{\alpha}^{2} + \Lambda_{i}^{2}}} 
\langle \alpha |\gamma_{3}\gamma_{5} \tau_{3} | \alpha\rangle\right]\,.
\label{eq:gavac} \end{equation}
The object in square brackets is the vacuum contribution to the axial 
charge \cite{Weigel:1999pc}. Similar 
calculations from the valence contribution give
\begin{equation}
\int d x \left[\mathsf{g}_{1}^{p}(x) -\mathsf{g}_{1}^{n}(x)\right]_{\rm v} 
= -\frac{N_c}{54} \left[1+{\rm sign}(\epsilon_{\rm v})\right]
\Big\langle {\rm v}\Big|\Vek{\tau}\cdot\Vek{\sigma}\Big|{\rm v}\Big\rangle= 
\frac{1}{6} \left[-\frac{N_C}{9}\left[1+{\rm sign}(\epsilon_{\rm v})\right]\langle {\rm v}| 
\gamma_{3}\gamma_{5}\tau_{3}| {\rm v}  \rangle \right]
\label{eq:gaval} \end{equation}
with the object in square brackets being the valence quark contribution to the axial
charge. This indeed verifies the Bjorken sum rule for the total axial charge.

In an analog, yet much more tedious, calculation the sum rules for the subleading
contributions in the $1/N_C$ expansion are also verified via level by level 
identities. Details may be found in Ref. \cite{TakyiThesis}. We would like to mention 
however, that the Adler sum rule \cite{Adler:1965ty}, which concerns a structure 
function from neutrino interactions and thus the exchange of a $W$ gauge boson (not 
considered here), measures the isospin of the nucleon. In that case the sum rule is 
not level by level; rather summing this integrated structure function over all levels 
reproduces the moment of inertia, Eq.~(\ref{mominert})~\cite{Weigel:1996ef}.

\section{Numerical results}
\label{sec:NR}

In this Section we present our numerical results for the structure functions. These 
results are obtained in a number of subsequent steps. First we construct the coordinate 
space eigenspinors of the self-consistent chiral soliton as described in Section \ref{sec:sol} 
for the parameters listed at the end of Section \ref{sec:model}. In the second step we 
evaluate the Fourier transform according to Eq.~(\ref{Fourier_trans}). Details of this 
transformation are provided in Appendix \ref{app:sol}. Essentially the spinors in
momentum space are combinations of spherical harmonic functions of the unit momentum
vector and momentum space radial functions that are Bessel transforms of the 
radial functions in the coordinate space spinors from Section \ref{sec:sol}.
In momentum space the spherical harmonic functions combine to the conserved 
grand spin just as do those in coordinate space. Hence we formally obtain matrix 
elements of operators as, for example $\Vek{\alpha}\cdot\hat{\Vek{p}}$, in the
very same way as the matrix elements of $\Vek{\alpha}\cdot\hat{\Vek{r}}$ in 
coordinate space. In the third step the momentum space radial functions are numerically 
integrated to produce the structure functions in the nucleon rest frame. In the next
step they are transformed to the infinite momentum frame \cite{Gamberg:1997qk} 
and subsequently the standard (perturbative QCD) evolution to the scale of the 
experimental data is performed to allow for a sensible comparison. We note that
this evolution brings into the game a new model parameter, the scale at which
the evolution commences. We take a single scale for all structure functions.

We test the outcome of our numerical simulations via the sum rules, that is, we compare 
the integrated functions with associated local quantity obtained from the coordinate 
space spinors. To gain acceptable agreement a very fine (equi-distant) grid for 
the radial variable in momentum space is required. Typically we take several 
thousand points on an interval between zero and ten times the physical cut-off, 
$\Lambda$. Needless to say that this consumes a large amount of CPU time and 
obtaining (in particular the subleading $1/N_C$ contributions to) the 
structure functions takes days or weeks on an ordinary desktop PC. Still, there 
are minor numerical inaccuracies as reflected by small oscillations of the structure 
functions around a central value at larger $x$ , {\it cf.} figures below.
Working in momentum space, rather than using the expansion coefficients 
$V_{\alpha\beta}$ introduced after Eq.~(\ref{diagh}) has, however, the advantage that no 
smearing \cite{Diakonov:1997vc} procedure is required. 

\subsection{Rest frame results}
\label{sec:RF}

\begin{figure}
\centerline{
	\includegraphics[width=15cm,height=5cm]{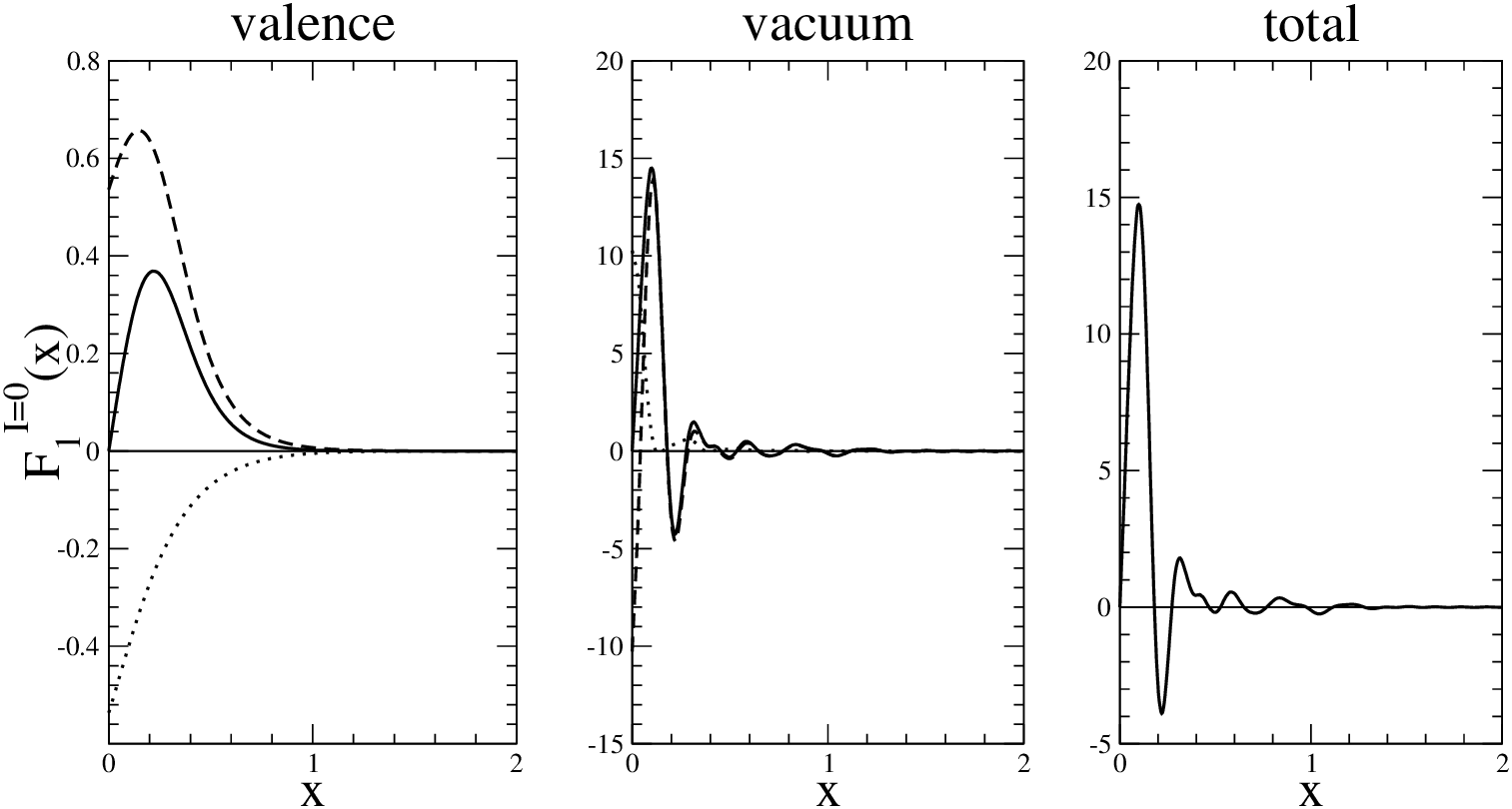}
}
\caption{Isoscalar unpolarized structure function in the nucleon rest
frame for a constituent quark mass of $400 {\rm MeV}$. Dashed and 
dotted lines refer to the positive and negative frequency contributions,
respectively.}
\label{f1a}
\end{figure}

\begin{figure}
\centerline{
        \includegraphics[width=15cm,height=5cm]{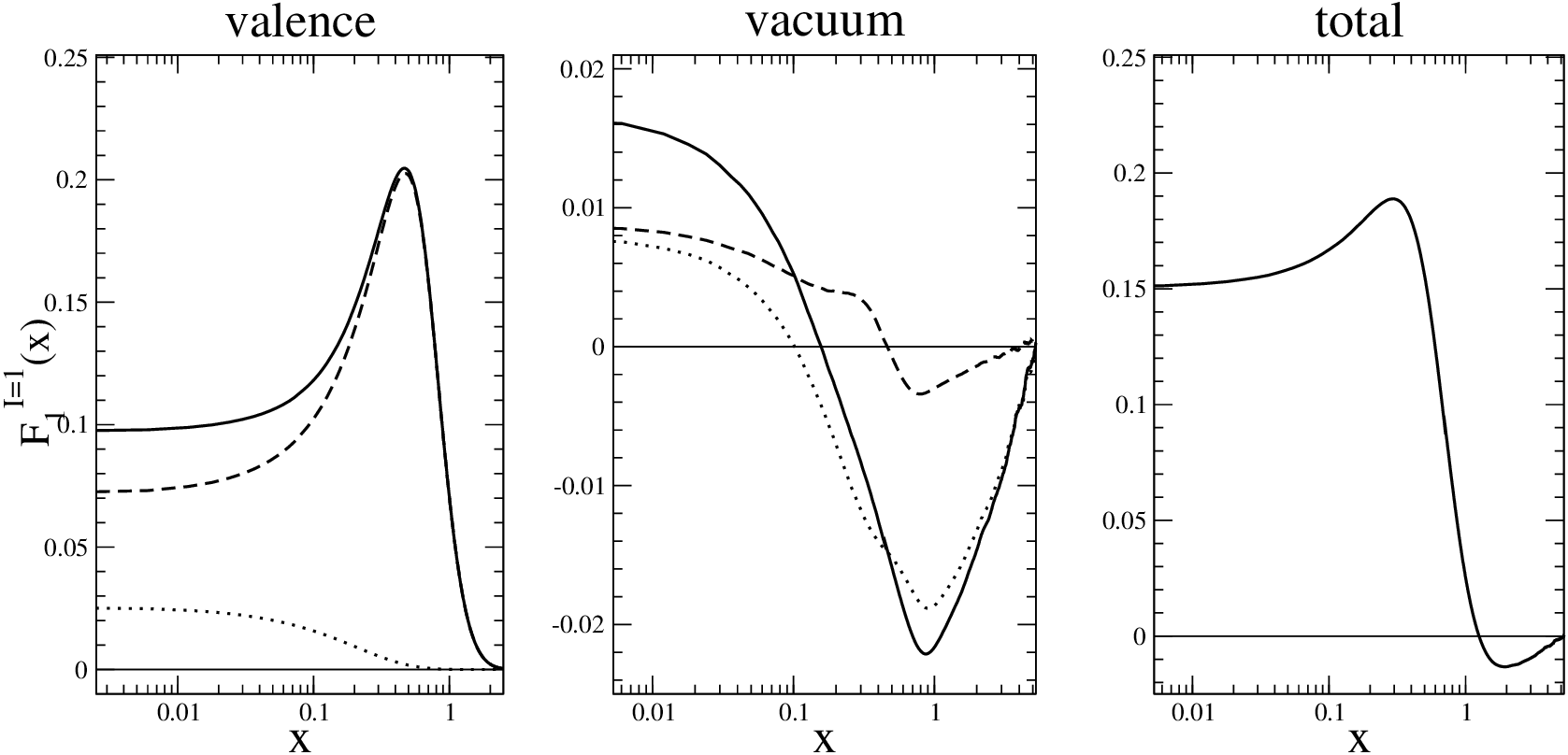}
}
\caption{Isovector unpolarized structure function in the nucleon rest frame for a 
constituent quark mass of $400 {\rm MeV}$. Note the logarithmic scale for the 
Bjorken variable. Dashed and dotted lines refer to the positive and negative 
frequency contributions, respectively.}
\label{f1b}
\end{figure}

In Figures \ref{f1a} and  \ref{f1b} we show typical results for the isoscalar and isovector 
components of the unpolarized structure functions respectively. In this case they have
been obtained using the constituent quark mass of $m=400 {\rm MeV}$. We separately show 
the contributions of the discrete valence level, those of the vacuum contributions and their 
sums (labeled as total). 
For the vacuum contribution we find the unexpected result that it dominates the 
valence counterpart. Mainly this originates from the (additional) subtraction
of the non-soliton piece mentioned after Eq.~(\ref{eq:f1s}). Without that 
subtraction we would not get a finite result, of course. Neither would the
momentum sum rule be fulfilled. However, this piece does not connect to the 
soliton rest frame and it is not clear at all whether or not transformation
of the Bjorken variable should be performed before taking the difference 
between the soliton and non-soliton isoscalar unpolarized structure functions.
Therefore we do not attach much relevance to this large vacuum contribution.
As expected, the vacuum contribution is sub-dominant for the isovector unpolarized 
structure function.

\begin{figure}
\centerline{
	\includegraphics[width=15cm,height=5cm]{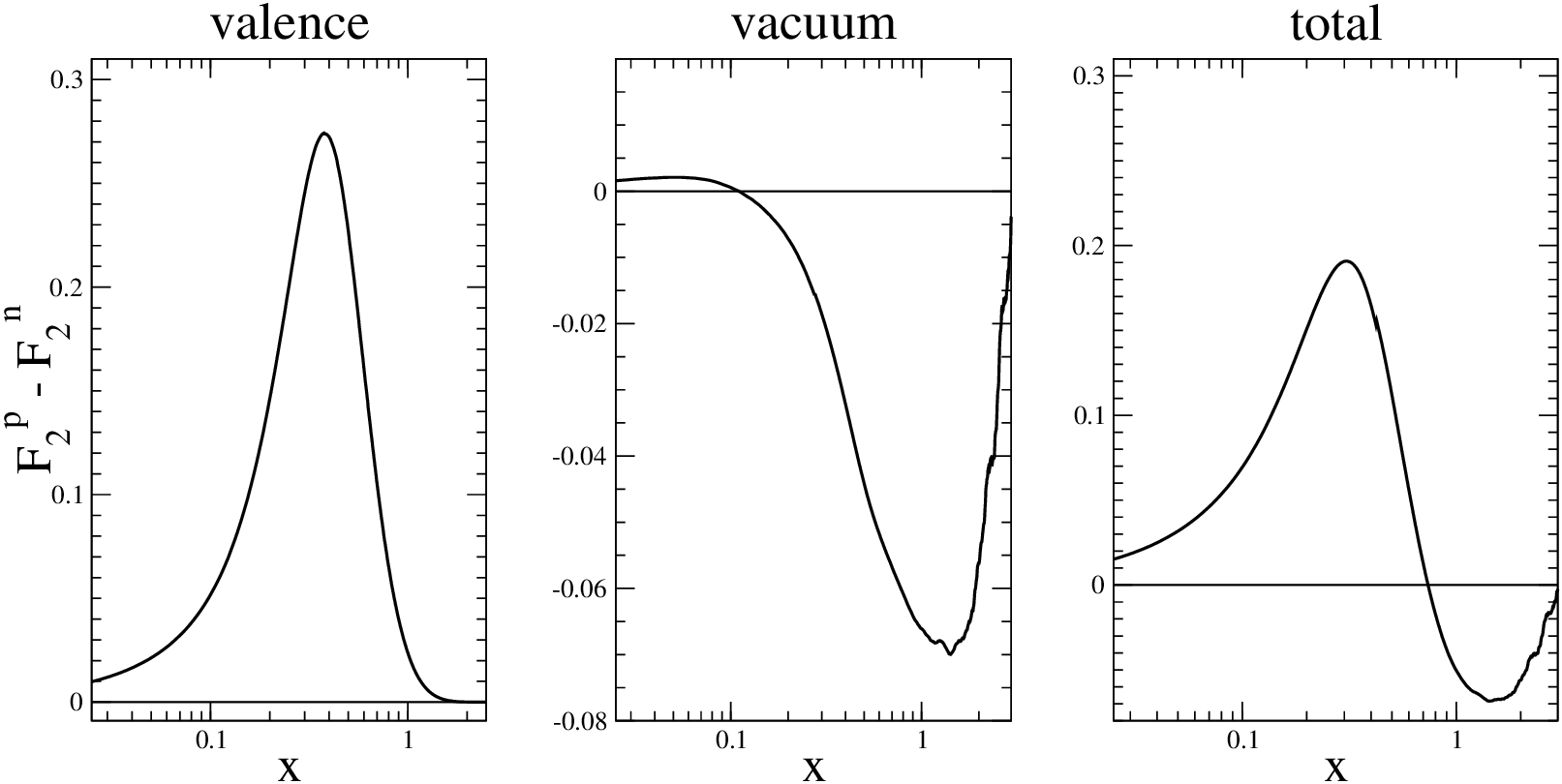}
}
\caption{Unpolarized structure function $F_{2}^{p}(x)-F_{2}^{n}(x)$ in the nucleon 
	rest frame for a constituent quark mass of $400 {\rm MeV}$.}
\label{f2}
\end{figure}
In Figure \ref{f2} we present the unpolarized structure function that enters the 
Gottfried sum rule, {\it i.e.} 
$F_{2}^{p}(x)-F_{2}^{n}(x) =2x \left[F_{1}^{p}(x)-F_{1}^{n}(x)\right]$ as the 
Callan-Gross relation holds in the soliton rest frame. The vacuum contribution turns 
slightly negative at large $x$ which persists when adding the dominating valence piece
to form the total contribution of this structure function. In table \ref{t1} we 
compare our results for the Gottfried sum rule, 
$\mathcal{S}_{G}= \int_{0}^{\infty}\frac{d x}{x}\,\left(F_{2}^{p}-F_{2}^{n}\right)$, for various 
constituent quark masses to the experimental data from the NM Collaboration \cite{Arneodo:1994sh}.
Under this integral, the vacuum part is even less significant as its positive and negative
parts compensate. In total, the agreement for the Gottfried sum rule is surprisingly good since 
usually chiral soliton models reproduce empirical data with 30\% accuracy \cite{Weigel:2008zz}.
\begin{table}
\caption{\label{t1}The Gottfried sum rule for various values of $m$. The subscripts 'v' and 's'
denote the valence and vacuum contributions, respectively. The third column contains their sums.}
\begin{center}
\begin{tabular}{|c| c| c| c|c|}
\hline 
$m\,[{\rm MeV}]$  & $[\mathcal{S}_{G}]_{{\rm v}}$  & $[\mathcal{S}_{G}]_{{\rm s}}$  
& $\mathcal{S}_{G}$ & empirical value \\
\hline
$400$  &  $0.214$   & $0.000156$  &  $0.214$ &  \\
$450$  &  $0.225$   & $0.000248$  &  $0.225$ &  $ 0.235 \pm 0.026 $ \cite{Arneodo:1994sh} \\
$500$  &  $0.236$   & $0.000356$  &  $0.237$ & \\
\hline 
\end{tabular}
\end{center}
\vskip-0.5cm
\end{table}

\begin{figure}
\centerline{
	\includegraphics[width=15cm,height=5cm]{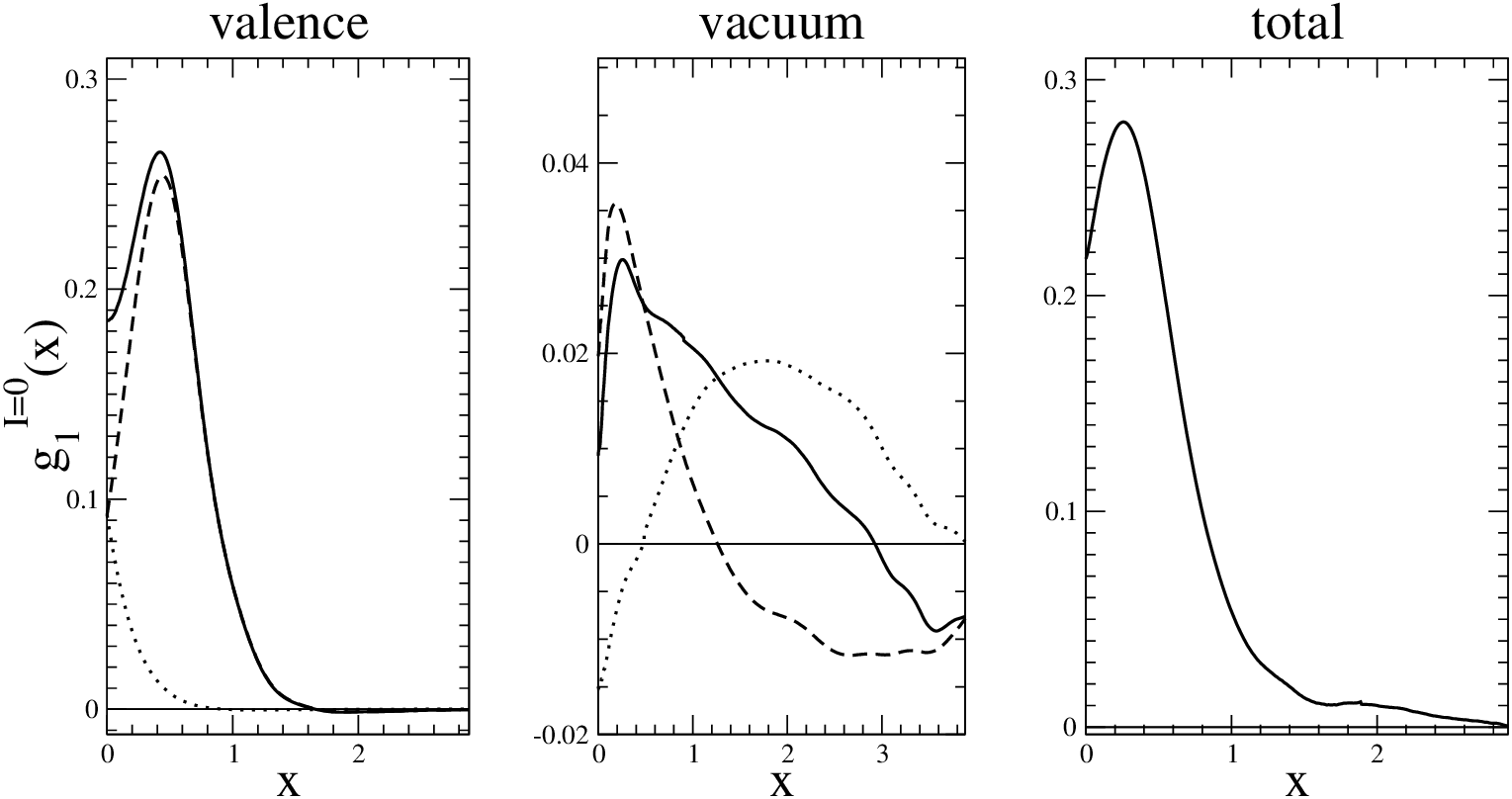}
}
\caption{Isoscalar longitudinal polarized structure functions in the 
nucleon rest frame for a constituent quark mass of $400 {\rm MeV}$. For the valence 
and vacuum contributions we separately display the positive (dashed) and 
negative (dotted) frequency contributions. The full lines are their sums. Note the
small vertical scale for the vacuum contribution.}
\label{f3s}
\end{figure}

\begin{figure}
\centerline{
        \includegraphics[width=15cm,height=5cm]{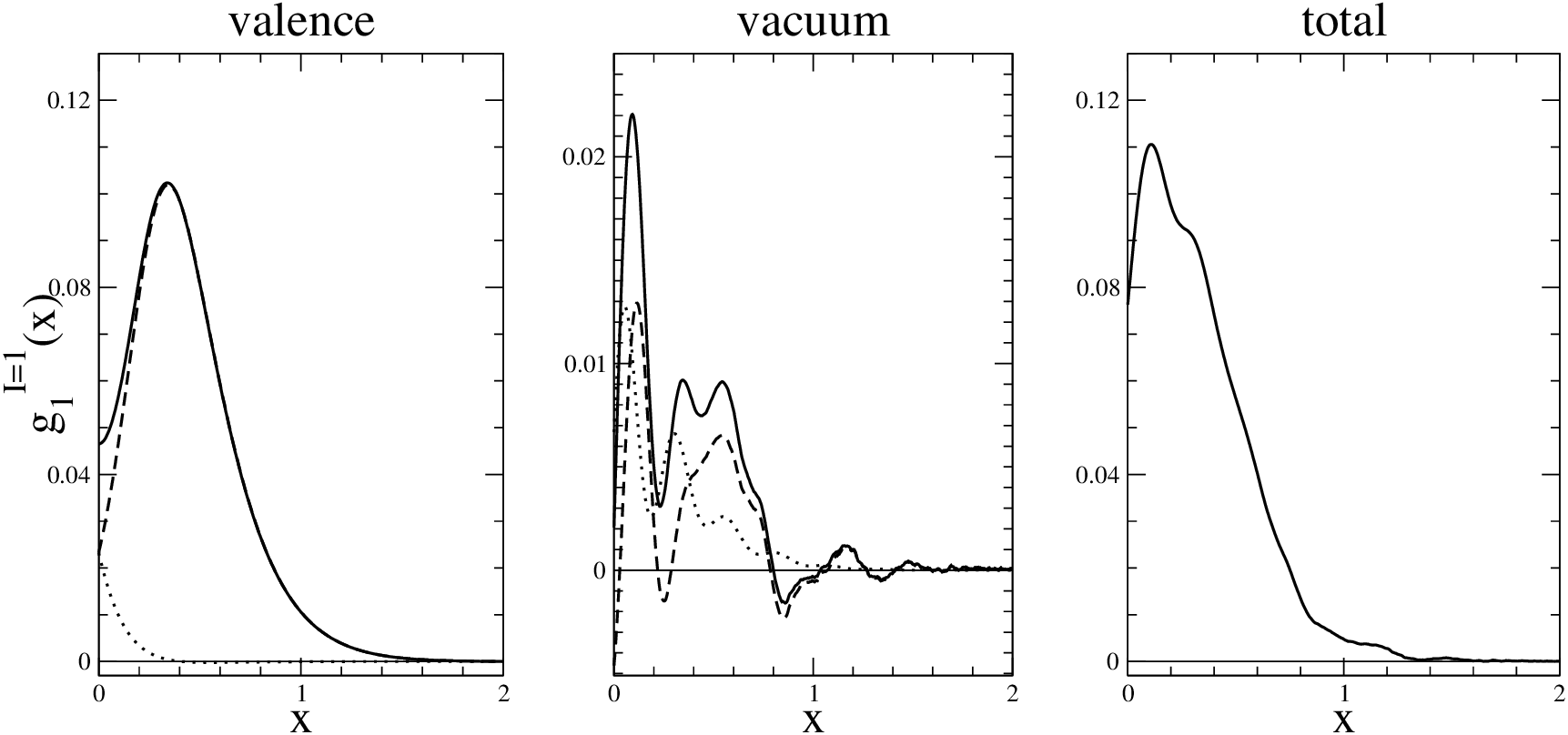}
}
\caption{Same as Figure \ref{f3s} for the isovector longitudinal polarized structure 
function.}
\label{f3v}
\end{figure}

In Figures \ref{f3s} and \ref{f3v} we show the isoscalar and isovector contributions to 
the longitudinal polarized structure functions $\mathsf{g}^{I=0,1}_1$, respectively. In both 
pictures we display the valence and vacuum contributions as well as their sums. Also 
the positive and negative frequency components of the valence and vacuum parts are shown. 
Obviously these structure functions are indeed dominated by their valence contributions and we thus 
a posteriori verify the {\it valence only approximation} adopted in Ref. \cite{Weigel:1996jh}.
The only exception is the isoscalar structure function $\mathsf{g}_{2}$ for which the valence 
contribution is small by its own due to large cancellations in 
$\mathsf{g}_2=\mathsf{g}_T-\mathsf{g}_1$. We also recognize some minor oscillations in the 
vacuum contributions at larger $x$. These occur as remnants of numerical inaccuracies.
We also remark that the present valence quark results do not exactly match those from 
Ref. \cite{Weigel:1996jh} in which a soliton profile from the proper-time regularization
scheme was employed.

We have computed the axial vector and singlet charges on one hand side via the 
respective sum rules, {\it i.e.} by integrating the structure functions $g^{I=1,0}_1(x)$
and on the other hand via the coordinate space matrix elements of $\tau_3\gamma_3\gamma_5$ 
and $\gamma_3\gamma_5$ as {\it e.g.} in Eqs.~(\ref{eq:gavac}) and~(\ref{eq:gaval}).
The comparison in table \ref{t2} serves as test for the numerical accuracy 
which works perfectly in the vector case while some minor discrepancies are observed for 
the axial singlet charge. This is understood as the latter is actually quartic in the
quark wave-functions (two of which are Fourier transformed) and it is also a double
sum over those wave-functions. So even tiny numerical errors  are amplified. 

As is typical for chiral soliton models, the axial vector charge falls short off the 
measured datum by about 30-40\% \cite{Weigel:2008zz}. It has been argued that this could be 
remedied by $1/N_C$ corrections arising from a particular handling of the collective 
coordinate quantization \cite{Wakamatsu:1993nq,Christov:1993ny}. However, these corrections do 
not emerge in the current approach and also lead to inconsistencies with PCAC \cite{Alkofer:1993pv}.
On the other hand, the predicted axial singlet charge, which is linked to the proton spin 
problem \cite{Aidala:2012mv}, is well within the errors of the empirical value.

\begin{figure}
\centerline{
	\includegraphics[width=15cm,height=5cm]{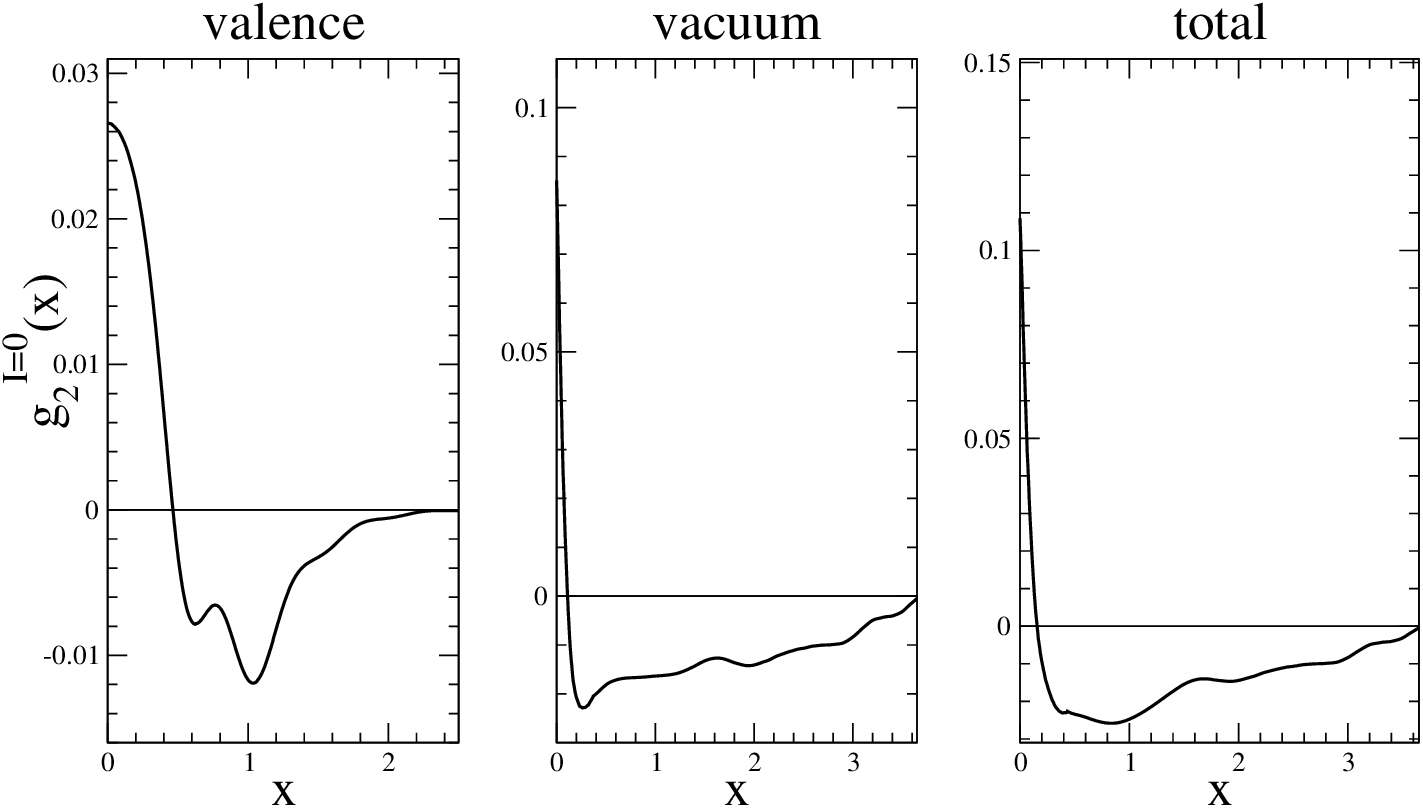}
}
\caption{Isoscalar structure function, $\mathsf{g}_2$, in the 
nucleon rest frame for a constituent quark mass of $400 {\rm MeV}$.}
\label{f4s}
\end{figure}

\begin{figure}
\centerline{
        \includegraphics[width=14cm,height=5cm]{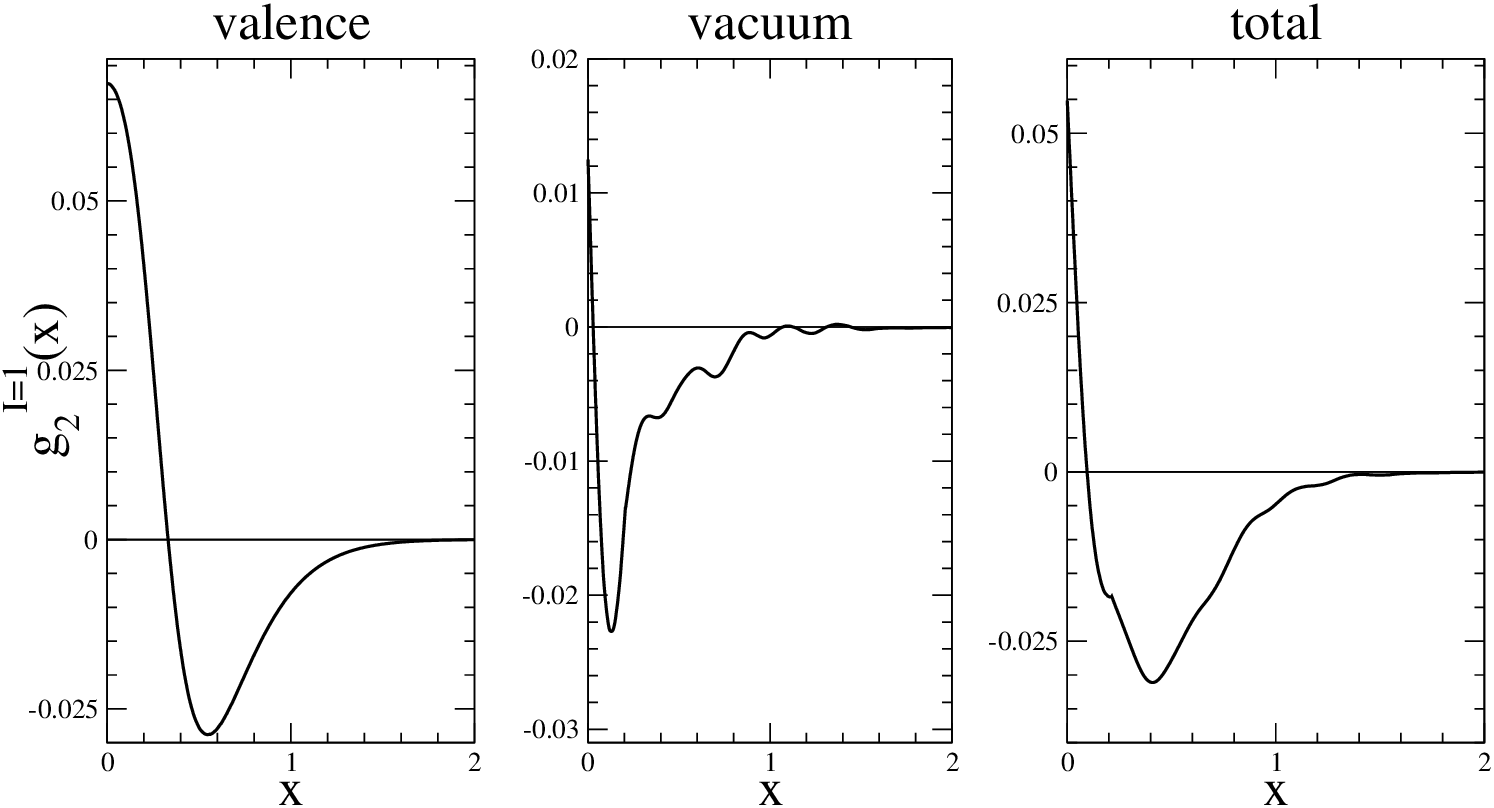}
}
\caption{Isovector polarized structure functions, $\mathsf{g}_2$, in the
nucleon rest frame for a constituent quark mass of $400 {\rm MeV}$.}
\label{f4v}
\end{figure}

\begin{table}[t]
\vskip-0.3cm
\caption{\label{t2}The axial-vector and -singlet charges for various values of the constituent 
quark mass $m$. Subscripts are as in table \ref{t1}. The data in parenthesis give the numerical
results as obtained from the coordinate space representation.}
\begin{center}
\begin{minipage}[b]{0.45\linewidth}
\begin{tabular}{|c| c| c| c| c|}
\hline 
$m\,[{\rm MeV}]$  & $[ g_{A}]_{{\rm v}}$ & $ [ g_{A}]_{{\rm s}}$  & $ g_{A}$
& empirical value  \\
\hline
$400$  &  $0.734$  & $0.065$ &  $0.799$  ($0.800$) & \\
$450$  &  $0.715$  & $0.051$ &  $0.766$  ($0.765$) & $1.2601\pm0.0025$ \cite{Barnett:1996hr}   \\
$500$  &  $0.704$  & $0.029$ &  $0.733$  ($0.733$) & \\
\hline 
\end{tabular}
\end{minipage}
\quad
\begin{minipage}[b]{0.45\linewidth}
\begin{tabular}{|c| c| c| c| c|}
\hline 
$m\,[{\rm MeV}]$  & $[ g_{A}^{0}]_{{\rm v}}$ & $[ g_{A}^{0}]_{\rm s}$ & $ g_{A}^{0}$ 
& empirical value  \\
\hline
$400$  &  $0.344$  & $0.0016$  &  $0.345$  ($0.350$) &  \\
$450$  &  $0.327$  & $0.0021$  &  $0.329$  ($0.332$) & $0.33 \pm 0.06$ \cite{Alexakhin:2006oza}  \\
$500$  &  $0.316$  & $0.0028$  &  $0.318$  ($0.323$) &  \\
\hline 
\end{tabular}
\end{minipage}
\end{center}
\vskip-0.2cm
\end{table}

\subsection{Projection and evolution}
\label{ssub:pande}

The soliton picture for baryons employs a localized field configuration which generally
breaks translational invariance. This causes the structure functions not to vanish when
$x>1$ as would be demanded kinematically. This effect is obvious in the above figures. 
We note that it is not limited to soliton models but is observed, {\it i.e.} in the bag
model as well \cite{Jaffe:1974nj}. Using light cone coordinates in the bag model in one 
space dimension a mapping of the structure functions from the localized field configuration 
was constructed that annihilated the structure functions for $x>1$~\cite{Jaffe:1980qx}. 
Guided by that construction a Lorentz boost was applied transforming the rest frame structure 
functions to the infinite momentum frame \cite{Gamberg:1997qk}
\begin{equation}
f_{IMF}(x) = \frac{\Theta \left(1-x\right)}{1-x} f_{RF} \left(-\ln (1-x) \right)\,,
\label{Infinite_momentum_frame} \end{equation} 
where $f_{RF}$ refers to any of the structure functions computed in Section \ref{sec:RF}.
In what follows we will omit the label $IMF$ for the boosted structure functions.

Even though we have adopted the high energy Bjorken limit in our kinematical analysis
of the Compton tensor, it must be emphasized that the NJL model is (at best) an approximation 
to QCD at the low mass scale, $\mu^{2}$ which is thus a hidden parameter in the approach.
To compare with experimental data that are taken at higher energy scales, $Q_{\rm exp}^2$ we adopt 
Altarelli-Parisi (or DGLAP) equations \cite{Gribov:1972ri,Altarelli:1977zs,Dokshitzer:1977sg}, 
to evolve the model structure functions accordingly. To be precise, we integrate
\begin{align}
f(x,t+\delta t)=f(x,t)+\delta t\frac{d f(x,t)}{d t}\,.\label{altarelli_procedure}
\end{align}
with $\displaystyle t = \ln \left( \frac{Q^2}{\Lambda^2_{QCD}} \right)$ from $Q^2=\mu^2$ to
$Q^2=Q_{\rm exp}^2$. The structure functions from Eq.~(\ref{Infinite_momentum_frame}) are 
the initial values and we tune $\mu^2$ for best fit at $Q_{\rm exp}^2$.

Since the isoscalar structure functions are associated with gluon type quantum numbers
they mix under the evolution. We take this into account under the assumption that
the gluon distributions vanish at $\mu^2$. We consider the leading order of the perturbative 
expansion sufficient to estimate the quality of our results. Then the evolution equations 
have the following structure 
\begin{align}
\frac{d f^{(I=1)}(x,t)}{d t}&=\frac{g_{QCD}(t)}{2 \pi}C_{R}(F)\int_{x}^{1} \frac{d y}{y}
P_{qq}(y) f^{(I=1)}\left(\frac{x}{y},t\right),\label{evol_integ_diff_equa_isovec} \\
\frac{d f^{(I=0)}(x,t)}{d t}&=\frac{g_{QCD}(t)}{2 \pi}C_{R}(F)\int_{x}^{1} \frac{d y}{y} 
\Biggl\{P_{qq}(y) f^{(I=0)}\left(\frac{x}{y},t\right)+P_{qg}(y) g\left(\frac{x}{y},t\right)
\Biggr\}, \label{evol_integ_diff_equa_isosca} \\
\frac{d g(x,t)}{d t}&=\frac{g_{QCD}(t)}{2 \pi}C_{R}(F)\int_{x}^{1}\frac{d y}{y} 
\Biggl\{P_{gg}(y) g\left(\frac{x}{y},t\right)+ P_{gq}(y)f^{(I=0)}\left(\frac{x}{y},t\right) 
\Biggr\}\,,\label{evol_integ_diff_equa_gluon}
\end{align}
where 
$\displaystyle C_{R}(F)= \frac{\left(N_{f}^{2}-1\right)}{2N_{f}}$ is the color
factor for $N_{f}$ flavors. Furthermore $\displaystyle g_{QCD}(t) = \frac{4\pi}{\beta_{0} t}$ with
$\displaystyle \beta_{0}=\frac{11}{3}N_{C}- \frac{2}{3}N_{f}$ is the leading order 
perturbative running coupling constant. Explicit expressions for the splitting functions 
$P_{qq},\ldots,P_{gg}$, taken from Ref. \cite{peskin1995introduction} are listed in 
Appendix \ref{app:split} for completeness. From the evolved isoscalar and isovector 
components we finally obtain the proton and neutron structure functions as sum
and difference
\begin{equation}
f^{(p,n)}(x,Q^2)=\frac{1}{2}\left[f^{I=1}(x,Q^2) \pm f^{I=0}(x,Q^2)\right]\,. 
\label{pro_neut_struc_func}
\end{equation}

We note that applying 
the perturbative QCD scheme to the model structure functions requires the identification
of model and QCD degrees of freedom even though there is no definite reason for doing
so other than the lack of any sensible alternative.

The second polarized structure function $\mathsf{g}_2$ contains subleading, twist three, 
elements that undergo a different evolution procedure that is also described in 
Appendix \ref{app:split}.

\subsection{Comparison with experiment}

As in previous calculations within the {\it valence only approximation}
\cite{Weigel:1996kw,Weigel:1996jh,Weigel:1998yy} we take $\mu^2=0.4{\rm GeV^2}$ 
as initial value in the evolution differential equations. Smaller values contradict 
the perturbative nature of the evolution procedure as $\frac{g_{QCD}(t)}{2 \pi}$ becomes 
sizable in view of $\Lambda_{QCD}^2=0.2 {\rm GeV^2}$. On the other hand, significantly
larger $\mu^2$ values worsen the agreement with experimental data.

\begin{figure}
\centerline{
\includegraphics[width=8cm,height=3.5cm]{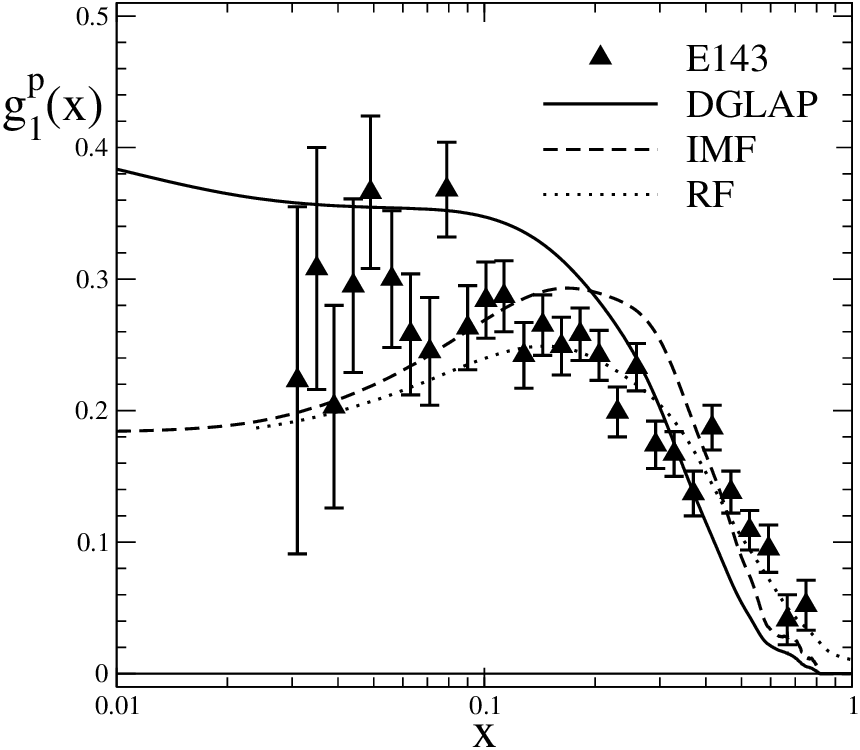}\hspace{1cm}
\includegraphics[width=8cm,height=3.5cm]{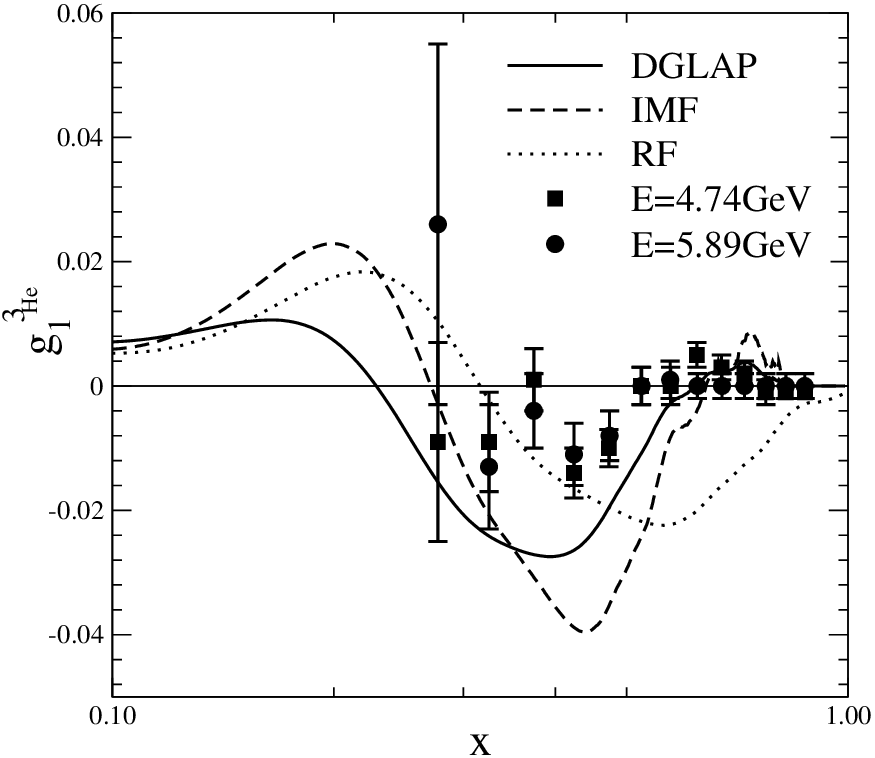}
}
\caption{Model prediction for the longitudinal polarized proton structure functions.  
Left panel: $\mathsf{g}_{1}^{p}(x)$ ; right panel: $\mathsf{g}_{1}^{^3{\rm He}}(x)$. 
These functions are ``DGLAP'' evolved from $\mu^{2}=\mathrm{0.4\,GeV^{2}}$ to 
$Q^{2}=\mathrm{3\,GeV^{2}}$ after projected to the infinite momentum frame ``IMF''. 
Data are from Refs. \cite{Abe:1994cp,Abe:1998wq} for the proton and 
from Ref. \cite{Flay:2016wie} for helium. In the latter case $E$ refers to the
electron energy.}
\label{f5}
\end{figure}
In the left panel of Figure \ref{f5} we show the numerical result for the polarized 
proton structure function $\mathsf{g}_1$ obtained from the evolution equation at 
$Q^{2}=3{\rm GeV^2}$. We compare our results to experimental results from the E143 
Collaboration \cite{Abe:1994cp,Abe:1998wq}. At small $x$ the model results are 
somewhat larger than the data, but definitely the gross features are predominantly
reproduced.

For the neutron data are available in terms of the helium structure 
function \cite{Flay:2016wie}\footnote{In Ref. \cite{Flay:2016wie} direct neutron 
data are only given as the ratio $\mathsf{g}_1^{n}(x)/F_1(x)$.}
\begin{equation}
\mathsf{g}_1^{^3{\rm He}}(x)\approx P_n \mathsf{g}_1^{n}(x)
+P_p \mathsf{g}_1^{p}(x) -0.014\left[\mathsf{g}_1^{p}(x)-4\mathsf{g}_1^{n}(x)\right]\,,
\label{eq:g1he}\end{equation}
with $P_n\approx 0.86$ and $P_p\approx-0.028$. From the right panel in Figure \ref{f5}
we see that our model results reproduce the main features of the data: small positive 
values at large $x$ turning negative at moderate $x$, though the minimum is more pronounced
by the model.

\begin{figure}
\bigskip
\centerline{
	\includegraphics[width=8cm,height=3.5cm]{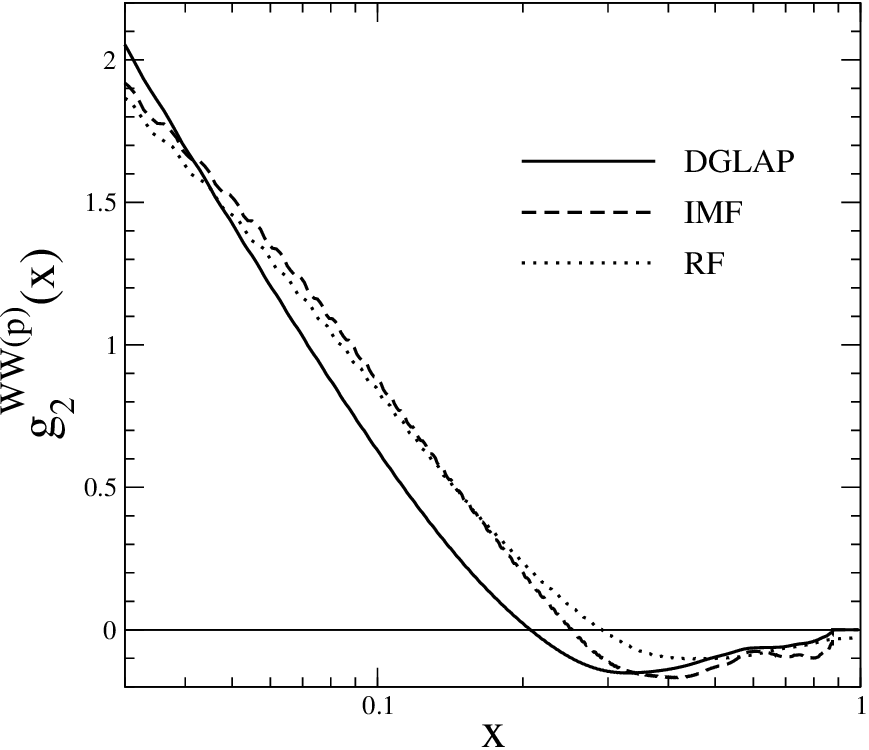}\hspace{1cm}
	\includegraphics[width=8cm,height=3.5cm]{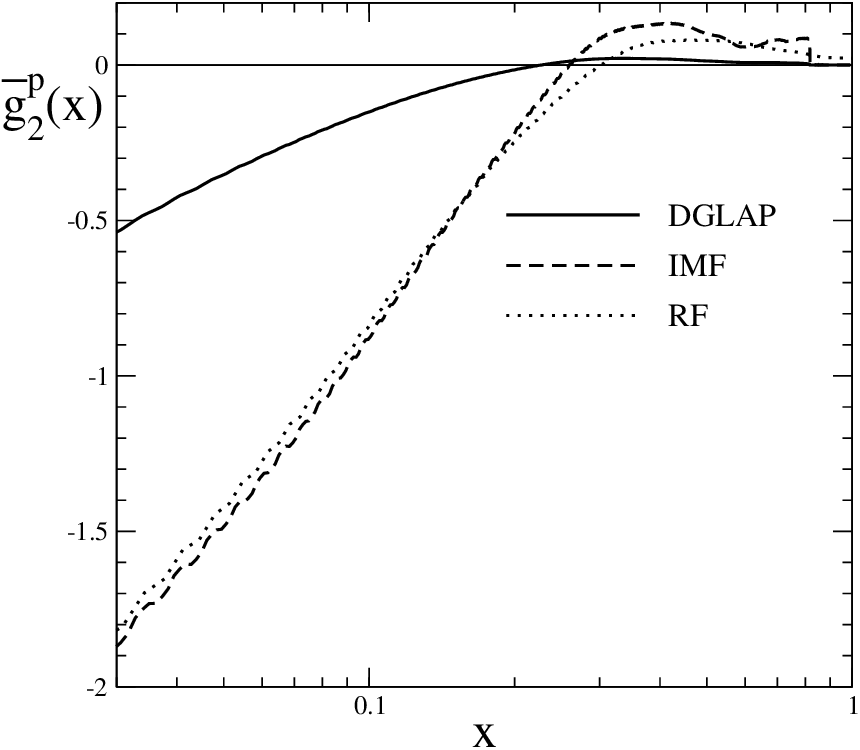}
}
\caption{Model prediction for the polarized proton structure functions 
$\mathsf{g}_{2}^{WW(p)}(x)$ (left panel) and
$\overline{\mathsf{g}}_{2}^{p}(x)$ (right panel) the are the twist-2 and -3 pieces
of $\mathsf{g}_{2}$. These functions are ``DGLAP'' evolved from 
$\mu^{2}=\mathrm{0.4\,GeV^{2}}$ to $Q^{2}=\mathrm{5\,GeV^{2}}$ after projected to 
the infinite momentum frame ``IMF''.}
\label{f6}
\end{figure}

Next we discuss the results for the structure function $\mathsf{g}_{2}(x)$. As discussed
in Appendix \ref{app:split} the twist-2 and -3 pieces must be disentangled within the 
evolution whose result is shown in Figure \ref{f6}. The effect of evolution is small 
for the twist-2 component but essential for the twist-3 counterpart.
\begin{figure}
\centerline{
\includegraphics[width=9cm,height=3.5cm]{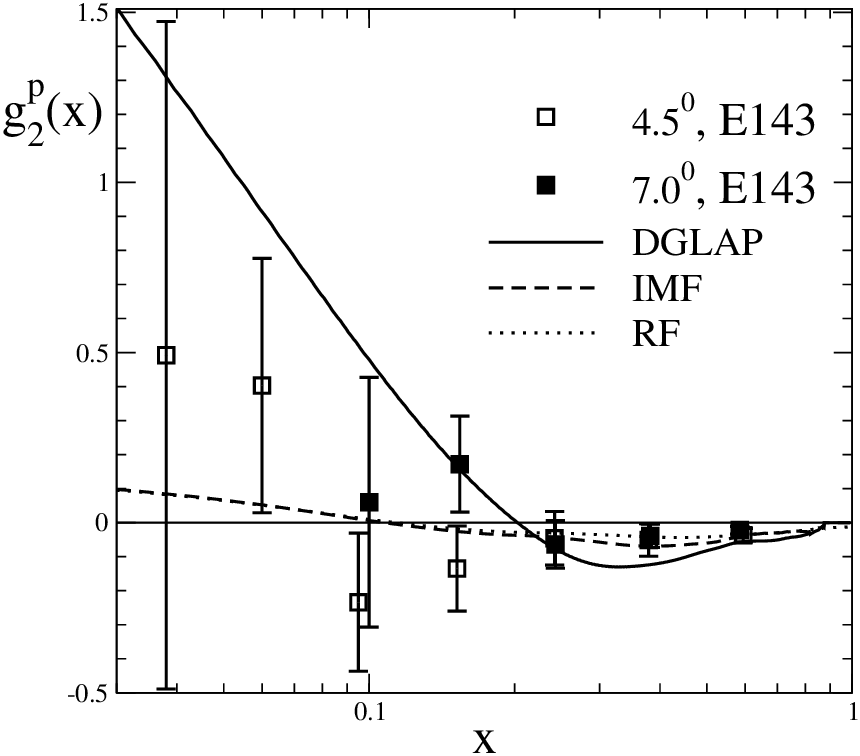}
}
\caption{Model prediction for the polarized proton structure functions 
$\mathsf{g}_{2}^{p}(x)$. This function is ``DGLAP'' evolved from $\mu^{2}=\mathrm{0.4\,GeV^{2}}$ to 
$Q^{2}=\mathrm{5\,GeV^{2}}$ after projected to the infinite momentum frame ``IMF''. Data is from Ref
\cite{Abe:1995dc}.}
\label{f7}
\end{figure}
When the end point of evolution is reached, the two components are combined to 
$\mathsf{g}_{2}^{p}(x,Q^2)$. We display the model prediction in Figure \ref{f7} and 
see that the data are well produced. This shows that the higher twist contributions 
cannot be neglected. This suggests that the higher twist contributions cannot be 
neglected.

Recently data were reported for the neutron twist-3 moment
\begin{equation}
d_2^n(Q^2)=3\int_0^1 dx\, x^2\, \overline{\mathsf{g}}^{n}_2(x,Q^2)
\label{eq:d2n}\end{equation}
at two different transferred momenta: $d_2^n(3.21{\rm GeV}^2)=-0.00421\pm0.00114$ and
$d_2^n(4.32{\rm GeV}^2)=-0.00035\pm0.00104$ \cite{Flay:2016wie}, where we added the listed
errors in quadrature. For $m=400{\rm MeV}$ the model calculation yields $-0.00426$ and
$-0.00409$, repsectively. While the lower $Q^2$ result nicely matches the observed value,
the higher one differs by about three standard deviations. This discrepancy as a function
of $Q^2$ indicates that the large $N_C$ approximation to evolve $\overline{\mathsf{g}}_2$ ({\it cf.}
Appendix \ref{app:split}) requires improvement.

\begin{figure}
\bigskip
\centerline{
\includegraphics[width=9cm,height=3.5cm]{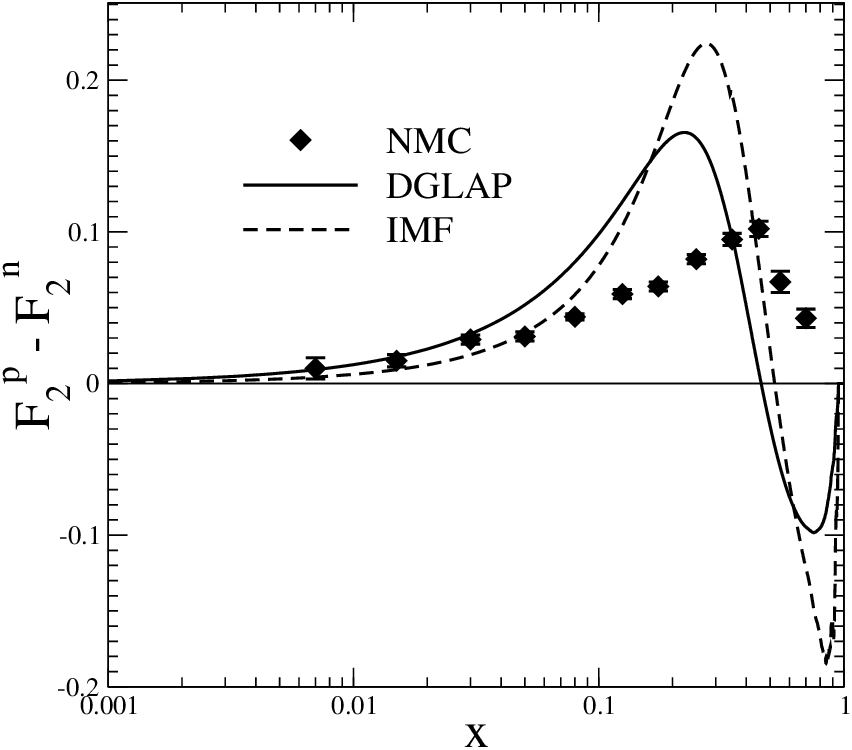}
}
\caption{Model prediction for the unpolarized structure function that enters the Gottfried sum rule. 
This function is ``DGLAP'' evolved from $\mu^{2}=\mathrm{0.4\,GeV^{2}}$ to $Q^{2}=\mathrm{4\,GeV^{2}}$ 
after transformation to the infinite momentum frame ``IMF''.}
\label{f8}
\end{figure}

Finally, in Figure \ref{f8} we display the unpolarized structure function that enters the Gottfried sum rule, 
{\it i.e.} $F_{2}^{p}(x)-F_{2}^{n}(x)$ using the evolution equation. Though the negative 
contribution from the Dirac vacuum ({\it cf.} Figure \ref{f1b}) around $x=1$ is tiny in 
the rest frame, it gets amplified when transforming to the infinite momentum frame by the 
factor $1/(1-x)$ in Eq.~(\ref{Infinite_momentum_frame}) thereby worsening the 
agreement with the experimental data collected by the NMC \cite{Arneodo:1994sh}.

\section{Conclusion}
\label{sec:concl}

We have presented the numerical simulation of nucleon structure functions within the 
NJL chiral soliton model. Central to this analysis has been the consistent implementation
of the regularized vacuum contributions that arise from all quark spinors being distorted
by the chiral soliton. Generally speaking, vacuum contributions should not be omitted
in any quark model as no expansion scheme suppresses them. This is even more the case
for the NJL model because the vacuum part significantly contributes to forming the soliton.

In our analysis we have only identified the symmetry currents of the model with those from QCD, 
not the quark distributions that are bilinear operators which are bilocal in the quark fields. 
Also, it is important to enforce the regularization on the action functional so that the 
regularization prescription for a given structure function is an unambiguous result. This 
increases the predictive power compared to previous similar studies that ''advocated'' an 
{\it ad hoc} regularization of quark distributions 
\cite{Diakonov:1997vc,Wakamatsu:1997en}\footnote{Schwinger's proper time regularization
scheme is popular in the context of the NJL chiral soliton \cite{Reinhardt:1989st}.
Ref.~\cite{Wakamatsu:1997en} explicitly states that its application to the quark 
distributions is not yet known.}.  A first principle regularization is particularly 
important when the sum rule for the structure function does not relate to coordinate space 
matrix elements of the quark fields. As an example we have seen that the isovector 
unpolarized structure functions are not subject to regularization (the explicit, 
lengthy formulas can be obtained from Ref. \cite{TakyiThesis}). The prediction for 
the corresponding, so-called Gottfried, sum rule decently matches the empirical value. 
This is a very favorable case for the {\it valence only approximation} as the vacuum 
contribution essentially integrates to zero due to an unexpected negative contribution at 
large $x$. For the isoscalar part we recognized that the subtraction of the zero-soliton 
vacuum contribution has a sizable effect at small $x$. The emergence of this 
contribution is somewhat surprising as it suggests that the zero-soliton vacuum has 
structure. Yet it is required for convergence as well as fulfilling the momentum sum 
rule. We emphasize that we observe acceptable agreement for polarized proton structure 
functions between our model results and the experimental data. For the polarized structure 
functions our numerically expensive computation indeed showed that the vacuum contribution
is sub-dominant, except maybe for the isoscalar part of $\mathsf{g}_2$ where the
valence part is tiny by itself. Nevertheless these results support the 
{\it valence only approximation} to a large extend.

In both, the unpolarized and polarized cases, the comparison with 
experiment required two additional operations on the model structure functions.
As the soliton is a localized field configuration, translational invariance is lost
and the rest frame structure functions must be Lorentz transformed to the infinite
momentum frame. In turn the results from that transformation are subject to the
perturbative QCD evolution scheme. This brings into the game the hidden parameter
at which to commence the evolution. We took that to be the same as in the 
{\it valence only approximation}.

Even though we have separated positive and negative frequency contributions to 
the structure functions we stop short of identifying them as (anti-)quark distributions
that parameterize semi-hard processes \cite{Gribov:1984tu}, like {\it e.g.} 
Drell-Yan \cite{Kenyon:1982tg}. The reason being that we avoid to identify model 
and QCD quark degrees of freedom at the model scale.
There are many other nucleon matrix elements of bilocal, bilinear quark operators 
for which experimental results or lattice data are available. Examples are chiral odd 
distributions \cite{Jaffe:1991kp,Jaffe:1991ra} or 
quasi-distributions \cite{Ji:2013dva,Alexandrou:2016jqi,Broniowski:2017wbr}. 
It is challenging to see whether quark distributions, or at least some 
of them, can also be formulated and computed with a first principle 
regularization scheme in the NJL chiral soliton model.

\acknowledgments
This project is is supported in part by the National Research Foundation of South Africa (NRF)
by grant~109497. I.~T. gratefully acknowledges a bursary from the {\it Stellenbosch Institute
for Advanced Studies} (STIAS).

\appendix

\section{Soliton matrix elements}
\label{app:sol}

The Dirac Hamiltonian $h$ of the hedgehog field configuration \eqref{hedgehog} commutes 
with the grand spin operator 
\begin{equation}
\Vek{G}=\Vek{J}+ \frac{\Vek{\tau}}{2} = \Vek{L} 
+ \frac{\Vek{\sigma}}{2} + \frac{\Vek{\tau}}{2},
\end{equation}
which is the operator sum of the total spin $\Vek{J}$ and the isospin $\Vek{\tau}/2$. 
The total spin is the operator sum of the orbital angular momentum $\Vek{L}$ and the 
intrinsic spin $\Vek{\sigma}/2$. Since $h$ preserves $\Vek{G}$, the eigenfunctions of 
the Dirac Hamiltonian are also eigenfunctions of $\Vek{G}$. The quantum numbers of 
$\Vek{G}$ are $\Vek{G}^{2}=G(G+1)$ and $G_{3}=M$. The respective  eigenfunctions are 
tensor spherical harmonics associated with the grand spin
\begin{equation}
\left[\mathcal{Y}_{LJGM}(\hat{\Vek{r}}) \right]_{si} =  \sum_{m, s_{3}, i_{3}, 
J_{3}} C^{GM}_{JJ_{3}, \frac{1}{2} i_{3}} C^{JJ_{3}}_{Lm, \frac{1}{2} s_{3}} 
\mathnormal{Y}_{Lm}(\hat{\Vek{r}}) \chi_{s}(s_{3}) \chi_{i}(i_{3})
\end{equation}
where $C^{GM}_{JJ_{3}, \frac{1}{2} i_{3}}$ and $C^{JJ_{3}}_{Lm, \frac{1}{2} s_{3}}$ 
are $SU(2)$ Clebsch-Gordon coefficients that describe the coupling of  $\chi_{s}$ and 
$\chi_{i}$, which are two components spinors and isospinors, respectively, and the spherical 
harmonic functions $Y_{Lm}$. 

For a prescribed profile function $\Theta (r)$ the numerical diagonalization 
of the Dirac Hamiltonian \eqref{hedgehog} produces the radial functions $g_{\alpha}^{(G,\pm;i)}$ 
and $f_{\alpha}^{(G,\pm;i)}(i=1,2)$ that feature in eight component spinors \cite{Kahana:1984be}
\begin{align}
\Psi_{\alpha}^{(G,+)}(\Vek{r}) = \begin{pmatrix}
\imu g_{\alpha}^{(G,+;1)}(r) \mathcal{Y}_{GG+\frac{1}{2}GM} (\hat{\Vek{r}}) \\
f_{\alpha}^{(G,+;1)}(r) \mathcal{Y}_{G+1G+\frac{1}{2}GM} (\hat{\Vek{r}})
\end{pmatrix}
+
\begin{pmatrix}
\imu g_{\alpha}^{(G,+;2)}(r) \mathcal{Y}_{GG-\frac{1}{2}GM} (\hat{\Vek{r}}) \\
-f_{\alpha}^{(G,+;2)}(r) \mathcal{Y}_{G-1G-\frac{1}{2}GM} (\hat{\Vek{r}})
\end{pmatrix}  \\
\Psi_{\alpha}^{(G,-)}(\Vek{r}) = \begin{pmatrix}
\imu g_{\alpha}^{(G,-;1)}(r) \mathcal{Y}_{G+1G+\frac{1}{2}GM} (\hat{\Vek{r}}) \\
-f_{\alpha}^{(G,-;1)}(r) \mathcal{Y}_{GG+\frac{1}{2}GM} (\hat{\Vek{r}})
\end{pmatrix}
+
\begin{pmatrix}
\imu g_{\alpha}^{(G,-;2)}(r) \mathcal{Y}_{G-1G-\frac{1}{2}GM} (\hat{\Vek{r}}) \\
f_{\alpha}^{(G,-;2)}(r) \mathcal{Y}_{GG-\frac{1}{2}GM} (\hat{\Vek{r}})
\end{pmatrix}.
\end{align}
Here, the second superscript $(\pm)$ denotes the intrinsic parity defined by the parity 
eigenvalue as $(-1)^{G}\times (\pm1)$. The radial functions are written as linear combinations 
of spherical Bessel functions that build the free spinors $\Psi_\alpha^{(0)}$. The order of 
these Bessel functions matches the angular momentum label (first subscript) of the multiplying 
$\mathcal{Y}$. The linear combination goes over momenta discretized by pertinent boundary 
conditions at a radius significantly larger than the extension of the profile function 
$\Theta (r)$. In Ref. \cite{Kahana:1984be} the condition that the radial function multiplying 
the $\mathcal{Y}$ with equal orbital angular momentum and grand spin indexes vanished at that 
large distance. In contrast, we impose that condition on the radial function of the upper 
component.  This avoids spurious contributions to the moment of inertia~\cite{Alkofer:1994ph}.
 
Writing 
\begin{equation}
{\rm e}^{\imu \Vek{p} \cdot \Vek{r}}= 4\pi \sum_{Lm} (\imu)^{L} j_{L}(pr) 
\mathnormal{Y}_{Lm}^{*} (\hat{\Vek{r}}) \mathnormal{Y}_{Lm}(\hat{\Vek{p}}).
\end{equation}
we find the Fourier transform, Eq.~(\ref{Fourier_trans}) of the spinors 
\begin{align}
\widetilde{\Psi}_{\alpha}^{(G,+)}(\Vek{p}) = (\imu)^{G+1} \begin{pmatrix}
\widetilde{g}_{\alpha}^{(G,+;1)}(p) \mathcal{Y}_{GG+\frac{1}{2}GM} (\hat{\Vek{p}}) \\
\widetilde{f}_{\alpha}^{(G,+;1)}(p) \mathcal{Y}_{G+1G+\frac{1}{2}GM} (\hat{\Vek{p}})
\end{pmatrix}
+
(\imu)^{G+1}\begin{pmatrix}
\widetilde{g}_{\alpha}^{(G,+;2)}(p) \mathcal{Y}_{GG-\frac{1}{2}GM} (\hat{\Vek{p}}) \\
\widetilde{f}_{\alpha}^{(G,+;2)}(p) \mathcal{Y}_{G-1G-\frac{1}{2}GM} (\hat{\Vek{p}})
\end{pmatrix}  \cr
\widetilde{\Psi}_{\alpha}^{(G,-)}(\Vek{p}) = -(\imu)^{G} \begin{pmatrix}
\widetilde{g}_{\alpha}^{(G,-;1)}(p) \mathcal{Y}_{G+1G+\frac{1}{2}GM} (\hat{\Vek{p}}) \\
\widetilde{f}_{\alpha}^{(G,-;1)}(p) \mathcal{Y}_{GG+\frac{1}{2}GM} (\hat{\Vek{p}})
\end{pmatrix}
+
(\imu)^{G}\begin{pmatrix}
\widetilde{g}_{\alpha}^{(G,-;2)}(p) \mathcal{Y}_{G-1G-\frac{1}{2}GM} (\hat{\Vek{p}}) \\
(\imu)^{G}\widetilde{f}_{\alpha}^{(G,-;2)}(p) \mathcal{Y}_{GG-\frac{1}{2}GM} (\hat{\Vek{p}})
\end{pmatrix}\,.
\end{align}
The radial functions in momentum space are the Fourier-Bessel transforms
\begin{equation}
\widetilde{\phi}_{\alpha} (p) = \int_0^\infty dr\, r^{2} j_{L_{\alpha}}(pr) \phi_{\alpha}(r)\,,
\label{eq:radial_momen}
\end{equation}
where $L_\alpha$ is the angular momentum associated with the coordinate space
radial wave-function $\phi_{\alpha}(r)$. Note that the grand spin spherical harmonic
functions in momentum space are constructed precisely as in coordinate space, just
that the argument is the momentum space solid angle. Note that the intrinsic parity is 
also conserved quantum number. 

The valence quark carries $G=0$, then only the components 
with $J=+1/2$ are allowed for the eigenspinor   
\begin{align}
\Psi_{\alpha}^{0,+} (\Vek{r})=\Psi_{\rm v}(\Vek{r}) = \begin{pmatrix}
\imu g_{\rm v}(r) \mathcal{Y}_{0,\frac{1}{2},0,0} (\hat{\Vek{r}}) \\
f_{\rm v}(r) \mathcal{Y}_{1,\frac{1}{2},0,0} (\hat{\Vek{r}})
\end{pmatrix}
\end{align}
here $g_{\rm v}(r) =g_{\alpha}^{(0,+;1)}(r)$ etc, are the particular eigenwave-functions. 
The cranking correction associated with the first order rotation \eqref{cranked_valence_level} 
dwells in the channel with $G=1$ and negative intrinsic parity 
\begin{align}
\Psi_{\alpha}^{(1,-)}(\Vek{r}) = \begin{pmatrix}
\imu g_{\alpha}^{(1)}(r) \mathcal{Y}_{2,\frac{3}{2},1,M} (\hat{\Vek{r}}) \\
-f_{\alpha}^{(1)}(r) \mathcal{Y}_{1,\frac{3}{2},1,M} (\hat{\Vek{r}})
\end{pmatrix}
+
\begin{pmatrix}
\imu g_{\alpha}^{(2)}(r) \mathcal{Y}_{0,\frac{1}{2},1,M} (\hat{\Vek{r}}) \\
f_{\alpha}^{(2)}(r) \mathcal{Y}_{1,\frac{1}{2},1,M} (\hat{\Vek{r}})
\end{pmatrix}, \label{cranked_function}
\end{align}
for convenience we have written $g_{\alpha}^{(1,-;1)}(r)$ as $g_{\alpha}^{(1)}(r)$ etc. 
Taking the Fourier transform of equation \eqref{cranked_valence_level} gives
\begin{equation}
\widetilde{\psi}_{\rm v} (\Vek{p})= \widetilde{\Psi}_{\rm v}(\Vek{p})+ \sum_{\alpha} 
\langle H_{\alpha} \rangle \widetilde{\Psi}_{\alpha}(\Vek{p}),
\label{fourier_transform_firs_order_rotation}
\end{equation}
where
\begin{equation}
\widetilde{\Psi}_{\rm v} (\Vek{p}) = \imu \begin{pmatrix}
\widetilde{g}_{\rm v}(p) \mathcal{Y}_{0,\frac{1}{2},0,0} (\hat{\Vek{p}}) \\
\widetilde{f}_{\rm v}(p) \mathcal{Y}_{1,\frac{1}{2},0,0} (\hat{\Vek{p}})
\end{pmatrix} \label{G0_momentum_represen}
\end{equation}
and 
\begin{align}
\widetilde{\Psi}_{\alpha} (\Vek{p}) = -\imu \begin{pmatrix}
\widetilde{g}_{\alpha}^{(1)}(p) \mathcal{Y}_{2,\frac{3}{2},1,M} (\hat{\Vek{p}}) -
\widetilde{g}_{\alpha}^{(2)}(p) \mathcal{Y}_{0,\frac{1}{2},1,M} (\hat{\Vek{p}})  \\
\widetilde{f}_{\alpha}^{(1)} (p) \mathcal{Y}_{1,\frac{3}{2},1,M} (\hat{\Vek{p}}) -
\widetilde{f}_{\alpha}^{(2)}(p) \mathcal{Y}_{1,\frac{1}{2},1,M} (\hat{\Vek{p}})
\end{pmatrix}. \label{G1_momentum_represen}
\end{align}
The ``matrix element'' $\langle H_{\alpha} \rangle$ arises from perturbatively treating
the collective rotation
\begin{equation}
\langle H_{\alpha} \rangle = \frac{1}{2} \frac{ \langle \alpha 
| \Vek{\tau} \cdot \Vek{\Omega}| {\rm v} \rangle}{\epsilon_{\rm v} - \epsilon_{\alpha}}.
\end{equation}

\section{Unpolarized Structure Functions at Leading Order}
\label{app:upolSF}

The level sums (over $\alpha$) as {\it e.g.} in Eq.~(\ref{vacuum_cont_F1}) concern the 
label of the radial function, grand spin ($G$) and its projection ($M$). As we average 
of the direction of the virtual photon \cite{Diakonov:1997vc}, the matrix elements are 
degenerate in $M$. This produces the extra factor $2G+1$ that we make explicit.

It is then straightforward to compute the matrix elements that appear 
in \eqref{unpolar_function}: 
\begin{align}
& \int d \Omega_{p} \widetilde{\Psi}_{\alpha}^{\dagger} (\Vek{p}) \widetilde{\Psi}_{\alpha} 
(\Vek{p}) \quad \text{and} \label{matrix_element_01} \\
& \int d \Omega_{p} \widetilde{\Psi}_{\alpha}^{\dagger} (\Vek{p}) \hat{\Vek{p}} \cdot 
\Vek{\alpha} \widetilde{\Psi}_{\alpha} (\Vek{p}) =  \int d \Omega_{p} 
\widetilde{\Psi}_{\alpha}^{\dagger} (\Vek{p}) \hat{\Vek{p}} \cdot \Vek{\sigma} 
\gamma_{5} \widetilde{\Psi}_{\alpha} (\Vek{p}) . \label{matrix_element_02}
\end{align}
The positive intrinsic parity of the matrix element of \eqref{matrix_element_01} 
is obtained as 
\begin{align}
\int d \Omega_{p} \widetilde{\Psi}_{\alpha}^{\dagger (G,+)} (\Vek{p}) 
\widetilde{\Psi}_{\alpha}^{(G,+)} (\Vek{p})  = (2G + 1) 
\Bigl( \widetilde{g}_{\alpha}^{(G,+;1)} (p)^{2} + \widetilde{f}_{\alpha}^{(G,+;1)} 
(p)^{2} + \widetilde{g}_{\alpha}^{(G,+;2)}(p)^{2} + \widetilde{f}_{\alpha}^{(G,+;2)} 
(p)^{2} \Bigr),
\end{align} 
and for the negative intrinsic parity as 
\begin{align}
\int d \Omega_{p}  \widetilde{\Psi}_{\alpha}^{\dagger(G,-)} (\Vek{p}) 
\widetilde{\Psi}_{\alpha}^{(G,-)} (\Vek{p})  = (2G+1) 
\Bigl( \widetilde{g}_{\alpha}^{(G,-;1)} (p)^{2} + \widetilde{f}_{\alpha}^{(G,-;1)} 
(p)^{2} + \widetilde{g}_{\alpha}^{(G,-;2)}(p)^{2} + \widetilde{f}_{\alpha}^{(G,-;2)} 
(p)^{2}\Bigr).
\end{align}
Here the overall factor $(2G+1)$ arises from summing the grand spin projection 
contained in $\sum_\alpha$. 
\begin{table}
\centering
\caption{Matrix elements $ \int d \Omega_{p} \mathcal{Y}_{L^{\prime}J^{\prime}GM}(\Vek{p})  
\hat{\Vek{p}} \cdot \Vek{\sigma} \mathcal{Y}_{LJGM}(\Vek{p})$.}
\label{t3}
\begin{tabular}{c c c c c c}
\hline
\multicolumn{2}{c}{$J^{\prime} = G - \frac{1}{2}$} &   
\multicolumn{2}{c}{ $J^{\prime}=G+\frac{1}{2}$}  &  &    \\    
$L^{\prime} = G-1$  & $L^{\prime} = G$ & $L^{\prime} = G$ 
& $L^{\prime} = G +1$ &  &  \\  
\hline
$0$    & $-1$ & $0$  &  $0$  &  $L=G-1$  & \\
&      &      &      &       & $J = G-\frac{1}{2}$\\
$-1$   & $0$  & $0$  &  $0$  &  $L=G$    & \\
$0$    & $0$  & $0$  &  $-1$ &  $L=G$    & \\
&      &      &      &       & $J=G+\frac{1}{2}$\\
$0$    & $0$  & $-1$ &  $0$  &  $L=G+1$  & \\
\hline
\end{tabular}	
\end{table}
From Table \eqref{t3} the positive intrinsic parity of the matrix element 
of \eqref{matrix_element_02} is obtained as 
\begin{align}
\int d \Omega_{p}\, \widetilde{\Psi}_{\alpha}^{\dagger (G,+)} (\Vek{p}) \hat{\Vek{p}} \cdot 
\Vek{\sigma} \gamma_{5} \widetilde{\Psi}_{\alpha}^{(G,+)} (\Vek{p})   =  -2(2G+1) 
\Bigl( \widetilde{g}_{\alpha}^{(G,+;1)}(p) \widetilde{f}_{\alpha}^{(G,+;1)}(p)  
+ \widetilde{g}_{\alpha}^{(G,+;2)}(p) \widetilde{f}_{\alpha}^{(G,+;2)}(p) \Bigr),
\end{align}
and that for the negative intrinsic parity as  
\begin{align*}
\int d \Omega_{p}\, \widetilde{\Psi}_{\alpha}^{\dagger (G,-)} (\Vek{p}) \hat{\Vek{p}} \cdot 
\Vek{\sigma} \gamma_{5} \widetilde{\Psi}_{\alpha}^{(G,-)} (\Vek{p})   = -2(2G+1) 
\Bigl( \widetilde{g}_{\alpha}^{(G,-;1)}(p) \widetilde{f}_{\alpha}^{(G,-;1)}(p)  
+\widetilde{g}_{\alpha}^{(G,-;2)}(p) \widetilde{f}_{\alpha}^{(G,-;2)}(p) \Bigr).
\end{align*}
The matrix element from the valence contribution \eqref{valence_cont_F1} is easily 
obtained, using the definition of the decomposition of the valence wave function 
\eqref{G0_momentum_represen}. They are given as 
\begin{align}
& \int d \Omega_{p} \widetilde{\Psi}_{\rm v}^{\dagger} (\Vek{p}) 
\widetilde{\Psi}_{\rm v}(\Vek{p})  = \widetilde{g}_{\rm v}(p)^{2} 
+ \widetilde{f}_{\rm v}(p)^{2} \quad \text{and} \\
& \int d \Omega_{p}  \widetilde{\Psi}_{\rm v}^{\dagger} (\Vek{p}) \hat{\Vek{p}} 
\cdot \Vek{\sigma} \gamma_{5} \widetilde{\Psi}_{\rm v}(\Vek{p})  = 
-2 \widetilde{g}_{\rm v}(p) \widetilde{f}_{\rm v}(p) 
\end{align}
at leading order $1/N_C$.

\section{Polarized Structure Functions at Leading Order}
\label{app:polSF}

Here we list the matrix elements that appear in the vacuum contribution of the 
polarized structure functions, Eqs.~\eqref{vac_cont_isovec_longpolar} and 
\eqref{vac_cont_isovec_transpolar}.
The matrix element to be considered are 
\begin{align}
& \int d \Omega_{p}  \widetilde{\Psi}_{\alpha}^{\dagger} (\Vek{p}) \hat{\Vek{p}} 
\cdot \Vek{\tau} \gamma_{5} \widetilde{\Psi}_{\alpha} (\Vek{p}) , 
\label{mat_elem_polar_01} \\
&\int d \Omega_{p}  \widetilde{\Psi}_{\alpha}^{\dagger} (\Vek{p}) \hat{\Vek{p}} \cdot 
\Vek{\tau} \hat{\Vek{p}} \cdot \Vek{\sigma} \widetilde{\Psi}_{\alpha} (\Vek{p}) 
\label{mat_elem_polar_02} \quad \text{and} \\
& \int d \Omega_{p} \widetilde{\Psi}_{\alpha}^{\dagger} (\Vek{p}) \Vek{\tau} 
\cdot \Vek{\sigma} \widetilde{\Psi}_{\alpha} (\Vek{p}).
\label{mat_elem_polar_03}
\end{align}
\begin{table}
\centering
\caption{Matrix elements $ \int d \Omega_{p} \mathcal{Y}_{L^{\prime} J^{\prime} 
GM}(\Vek{p}) \hat{\Vek{p}} \cdot \Vek{\tau} \mathcal{Y}_{LJGM}(\Vek{p}) $. 
The overall factor $1/(2G+1)$} needs to be multiplied.
\label{t4}
\begin{tabular}{c c c c c c}
\hline
\multicolumn{2}{c}{$J^{\prime} = G - \frac{1}{2}$} &   
\multicolumn{2}{c}{ $J^{\prime}=G+\frac{1}{2}$}  &  &   \\    
$L^{\prime} = G-1$  & $L^{\prime} = G$ & $L^{\prime} = G$ 
& $L^{\prime} = G +1$ &  &  \\  
\hline
$0$   & $-1$  &    $-2\sqrt{G(G+1)}$  &   $0$   &  $L=G-1$  & \\
&  &  &  &   & $J = G-\frac{1}{2}$\\
$-1$  &  $0$ & $0$  &  $-2\sqrt{G(G+1)}$ &  $L=G$    & \\
$-2\sqrt{G(G+1)}$  &   $0$ & $0$ & $1$  &  $L=G$    & \\
& &  & &    & $J=G+\frac{1}{2}$\\
$0$ & $-2\sqrt{G(G+1)}$ & $1$ & $0$  &  $L=G+1$  & \\
\hline
\end{tabular}
\end{table}
The matrix element \eqref{mat_elem_polar_01} is computed from the matrix elements from 
Table \ref{t4}: the positive intrinsic parity is obtained as
\begin{align}
\int d \Omega_{p}\, \widetilde{\Psi}_{\alpha}^{\dagger (G,+)} (\Vek{p}) \hat{\Vek{p}}
\cdot \Vek{\tau}  \gamma_{5} \widetilde{\Psi}_{\alpha}^{(G,+)} (\Vek{p}) 
& = 2\left(\widetilde{g}_{\alpha}^{(G,+;1)} (p) \widetilde{f}_{\alpha}^{(G,+;1)} (p) -  
\widetilde{g}_{\alpha}^{(G,+;2)}(p) \widetilde{f}_{\alpha}^{(G,+;2)}(p) \right) \cr
& - 4 \sqrt{G(G+1)} \Bigl( \widetilde{g}_{\alpha}^{(G,+;1)}(p) \widetilde{f}_{\alpha}^{(G,+;2)}(p)  
+  \widetilde{g}_{\alpha}^{(G,+;2)}(p)\widetilde{f}_{\alpha}^{(G,+;1)} (p) \Bigr)\,,
\end{align}
and that for negative intrinsic parity as
\begin{align}
\int d \Omega_{p}\,  \widetilde{\Psi}_{\alpha}^{\dagger (G,-)} (\Vek{p}) \hat{\Vek{p}} 
\cdot \Vek{\tau} \gamma_{5} \widetilde{\Psi}_{\alpha}^{(G,-)} (\Vek{p}) 
& = 2\left( \widetilde{g}_{\alpha}^{(G,-;1)} (p) \widetilde{f}_{\alpha}^{(G,-;1)} (p) - 
 \widetilde{g}_{\alpha}^{(G,-;2)}(p) \widetilde{f}_{\alpha}^{(G,-;2)}(p) \right) \cr
& + 4 \sqrt{G(G+1)} \Bigl( \widetilde{g}_{\alpha}^{(G,-;1)}(p) \widetilde{f}_{\alpha}^{(G,-;2)}(p)  
+  \widetilde{g}_{\alpha}^{(G,-;2)}(p)\widetilde{f}_{\alpha}^{(G,-;1)} (p) \Bigr)\,.
\end{align}
Also the matrix element \eqref{mat_elem_polar_02} is computed from the matrix elements 
from Table \ref{t5}:
\begin{table}
\centering
\caption{Matrix elements $ \int d \Omega_{p} \mathcal{Y}_{L^{\prime} J^{\prime} GM}(\Vek{p}) 
\hat{\Vek{p}} \cdot \Vek{\tau} \hat{\Vek{p}}\cdot \Vek{\sigma} \mathcal{Y}_{LJGM}(\Vek{p}) $. 
The overall factor $1/(2G+1)$} needs to be multiplied.
\label{t5}
\begin{tabular}{c c c c c c}
\hline
\multicolumn{2}{c}{$J^{\prime} = G - \frac{1}{2}$} &   
\multicolumn{2}{c}{ $J^{\prime}=G+\frac{1}{2}$}  &  &    \\    
$L^{\prime} = G-1$  & $L^{\prime} = G$ & $L^{\prime} = G$ 
& $L^{\prime} = G +1$ &  &  \\  
\hline
$1$  &    $0$  &    $0$   &  $2\sqrt{G(G+1)}$   &  $L=G-1$  & \\
&   &  &   &    & $J = G-\frac{1}{2}$\\
$0$   &  $1$ & $2\sqrt{G(G+1)}$ &  $0$ &  $L=G$    & \\
$0$  & $2\sqrt{G(G+1)}$ & $-1$  &  $0$  &  $L=G$    & \\
&    &    &     &   & $J=G+\frac{1}{2}$\\
$2\sqrt{G(G+1)}$    &    $0$  &    $0$ &  $-1$  &  $L=G+1$  & \\
\hline
\end{tabular}
\end{table}
the positive intrinsic parity contribution is given as
\begin{align}
\int d \Omega_{p}\, \widetilde{\Psi}_{\alpha}^{\dagger (G,+)} (\Vek{p}) \hat{\Vek{p}}
\cdot \Vek{\tau} \hat{\Vek{p}} \cdot \Vek{\sigma} \widetilde{\Psi}_{\alpha}^{(G,+)} (\Vek{p}) 
& = \Bigl(- \widetilde{g}_{\alpha}^{(G,+;1)} (p)^{2} - \widetilde{f}_{\alpha}^{(G,+;1)} (p)^{2}  
+\widetilde{g}_{\alpha}^{(G,+;2)} (p)^{2} + \widetilde{f}_{\alpha}^{(G,+;2)} (p)^{2} \Bigr)\cr
&  + 4\sqrt{G(G+1)} \Bigl( \widetilde{g}_{\alpha}^{(G,+;1)}(p) \widetilde{g}_{\alpha}^{(G,+;2)}(p)  
+ \widetilde{f}_{\alpha}^{(G,+;1)}(p) \widetilde{f}_{\alpha}^{(G,+;2)} (p) \Bigr),
\end{align}
and that for the negative intrinsic parity as
\begin{align}
\int d \Omega_{p}\,  \widetilde{\Psi}_{\alpha}^{\dagger (G,-)} (\Vek{p}) \hat{\Vek{p}}
\cdot \Vek{\tau} \hat{\Vek{p}} \cdot \Vek{\sigma} \widetilde{\Psi}_{\alpha}^{(G,-)} (\Vek{p}) 
& = \Bigl( -\widetilde{g}_{\alpha}^{(G,-;1)} (p)^{2} - \widetilde{f}_{\alpha}^{(G,-;1)} (p)^{2}  
+ \widetilde{g}_{\alpha}^{(G,-;2)} (p)^{2} + \widetilde{f}_{\alpha}^{(G,-;2)} (p)^{2} \Bigr) \cr
&  - 4\sqrt{G(G+1)} \Bigl( \widetilde{g}_{\alpha}^{(G,-;1)}(p) 
\widetilde{g}_{\alpha}^{(G,-;2)}(p)  + \widetilde{f}_{\alpha}^{(G,-;1)}(p) 
\widetilde{f}_{\alpha}^{(G,-;2)} (p) \Bigr). 
\end{align}
Furthermore the matrix element \eqref{mat_elem_polar_03} is computed from the matrix 
elements from Table \ref{t6}:
\begin{table}
\centering
\caption{Matrix elements $ \int d \Omega_{p} \mathcal{Y}_{L^{\prime} J^{\prime} GM} (\Vek{p}) 
\Vek{\tau} \cdot \Vek{\sigma} \mathcal{Y}_{LJGM}(\Vek{p})$. 
The overall factor $1/(2G+1)$} needs to be multiplied.
\label{t6}
\begin{tabular}{c c c c c c}
\hline
\multicolumn{2}{c}{$J^{\prime} = G - \frac{1}{2}$} &   
\multicolumn{2}{c}{ $J^{\prime}=G+\frac{1}{2}$}  &  &    \\    
$L^{\prime} = G-1$  & $L^{\prime} = G$ & $L^{\prime} = G$ 
& $L^{\prime} = G +1$ &  &  \\  
\hline
$2G+1$ &  $0$  & $0$  & $0$ &  $L=G-1$  & \\
&   &   &   &   & $J = G-\frac{1}{2}$\\
$0$  &  $-(2G-1)$  &  $4\sqrt{G(G+1)}$ &   $0$   &  $L=G$   & \\
$0$  & $4\sqrt{G(G+1)}$  &  $-(2G+3)$  &  $0$   &  $L=G$    & \\
&  &  &   &   & $J=G+\frac{1}{2}$\\
$0$  & $0$  & $0$ &  $2G+1$ &  $L=G+1$  & \\
\hline
\end{tabular}
\end{table}
the positive intrinsic parity contribution becomes
\begin{align}
\int d \Omega_{p}\,  \widetilde{\Psi}_{\alpha}^{\dagger (G,+)} (\Vek{p}) 
\Vek{\tau} \cdot \Vek{\sigma} \widetilde{\Psi}_{\alpha}^{(G,+)} (\Vek{p}) 
& = (2G+1) \left( \widetilde{f}_{\alpha}^{(G,+;1)} (p)^{2} 
+ \widetilde{f}_{\alpha}^{(G,+;2)}(p)^{2} \right) - (2G+3) 
\widetilde{g}_{\alpha}^{(G,+;1)}(p)^{2} \cr 
& + (2G-1) \widetilde{g}_{\alpha}^{(G,+;2)} (p)^{2}  + 8 \sqrt{G(G+1)} 
\widetilde{g}_{\alpha}^{(G,+;1)}(p) \widetilde{g}_{\alpha}^{(G,+;2)}(p)\,,
\end{align}
and for negative intrinsic parity becomes
\begin{align}
\int d \Omega_{p}\,  \widetilde{\Psi}_{\alpha}^{\dagger (G,-)} (\Vek{p}) 
\Vek{\tau} \cdot \Vek{\sigma} \widetilde{\Psi}_{\alpha}^{(G,-)} (\Vek{p}) 
& = (2G+1) \left( \widetilde{g}_{\alpha}^{(G,-;1)} (p)^{2} + 
\widetilde{g}_{\alpha}^{(G,-;2)}(p)^{2} \right) - (2G+3) 
\widetilde{f}_{\alpha}^{(G,-;1)}(p)^{2} \cr
& - (2G-1) \widetilde{f}_{\alpha}^{(G,-;2)} (p)^{2}  - 8 \sqrt{G(G+1)} 
\widetilde{f}_{\alpha}^{(G,-;1)}(p) \widetilde{f}_{\alpha}^{(G,-;2)} (p)\,.
\end{align}

\section{Splitting functions}
\label{app:split}

Here we list the  splitting functions used in Eqs.~(\ref{evol_integ_diff_equa_isovec}), 
(\ref{evol_integ_diff_equa_isosca}) and~(\ref{evol_integ_diff_equa_gluon}). They are 
different for the isovector, isosinglet and gluon contributions and are 
given as \cite{Altarelli:1977zs,Altarelli:1993np} 
\begin{align}
P_{qq}(z)&=\frac{1+z^{2}}{\left(1-z\right)_{+}} ,\cr
P_{qg}(z)&=\frac{1}{2C_{R}(F)}\left[z^{2} + (1-z)^{2}\right], \cr
P_{gq}(z)&=\frac{1+(1-z)^{2}}{z}, \cr
P_{gg}(z)&=2\Big[\frac{1-z}{z}+\frac{z}{(1-z)_{+}}+z(1-z)\Big].\label{splitting_functions}
\end{align}
They determine the probability for the parton $m$ to emit a parton $n$ such that the 
momentum of the parton $m$ is reduced by the fraction $z$. The regularized function 
$(1-z)^{-1}_{+}$ is defined under the integral by \cite{Altarelli:1977zs}
\begin{equation}
\int_{0}^{1} d z\,\frac{f(z)}{\left(1-z\right)_{+}}=\int_{0}^{1} d z\,\frac{f(z) -f(1)}{1-z}. \label{reg_plus_prescrip}
\end{equation}
In the above $\displaystyle C_{R}(F)= \frac{\left(N_{f}^{2}-1\right)}{2N_{f}}$ is the color 
factor for $N_{f}$ flavors. Also the running coupling constant in the leading order is given 
by $\displaystyle g_{QCD}(t) = \frac{4\pi}{\beta_{0} t}$ with 
$\displaystyle \beta_{0}=\frac{11}{3}N_{c}- \frac{2}{3}N_{f}$ being the coefficient of the leading 
term of the QCD $\beta$-function. Using the ``$+$ '' prescription, the evolution equations 
for the isovector, isosinglet and gluon contributions become \cite{Altarelli:1993np}
\begin{align}
\frac{d f^{(I=1)}(x,t)}{d t}&=\frac{2C_{R}(F)}{9t}\Bigg\{ \int_{x}^{1} 
\frac{d y}{y}\left(\frac{1+y^{2}}{1-y}\right)\Bigg[\frac{1}{y} f^{(I=1)}
\left(\frac{x}{y},t\right) - f^{(I=1)}(x,t) \Bigg] \cr 
& + \Bigg[ x + \frac{x^{2}}{2} + 2\ln \left(1-x\right) \Bigg] 
f^{(I=1)}(x,t) \Bigg\},\cr 
\frac{d f^{(I=0)}(x,t)}{d t}&=\frac{2C_{R}(F)}{9t}\Bigg\{\int_{x}^{1} 
\frac{d y}{y}\Bigg(\left(\frac{1+y^{2}}{1-y} \right)\Bigg[\frac{1}{y} 
f^{(I=0)}\left(\frac{x}{y},t\right) - f^{(I=0)}(x,t) \Bigg] \cr 
&+ \frac{3}{4}\left(y^{2}+(1-y^{2})\right) \frac{g(x,t)}{y} \Bigg) 
+ \Bigg[ x + \frac{x^{2}}{2} + 2\ln \left(1-x\right) \Bigg] 
f^{I=0}(x,t) \Bigg\}, \cr 
\frac{d g(x,t)}{d t}&=\frac{2C_{R}(F)}{9t}\Bigg\{\int_{x}^{1} 
\frac{d y}{y}\Bigg(\left(\frac{1+(1-y)^{2}}{y} \right)f^{(I=0)}
\left(\frac{x}{y},t\right) \cr 
&+\frac{9}{2}\left(\frac{1-y}{y} + y^{2}(1-y)\right)
\frac{ g\left(\frac{x}{y},t\right)}{y} + \frac{9}{2} 
\frac{g\left(\frac{x}{y},t\right)-g(x,t)}{1-y} \Bigg)  
+\Bigg[\frac{3}{2} + \frac{9}{2}\ln \left(1-x\right) \Bigg] g(x,t) \Bigg\}.
\label{app:eveqn}
\end{align}
Now, since our NJL model calculations do not account for any gluon content in the 
nucleon, we assume in our numerical calculations that, at the initial boundary scale 
$\mu^{2}$, the gluon content, $f(x,t_0)=0$ for both the polarized and unpolarized 
structure functions.

Unlike the polarized spin structure function $\mathsf{g}_{1}(x)$ of the nucleon and the unpolarized 
structure functions, the nucleon's second polarized spin structure function $\mathsf{g}_{2}$ involves 
contributions from quark-gluon iterations and quark masses \cite{Jaffe:1989xx,Ji:1990br,Jaffe:1990qh}. 
According to the standard operator product expansion analysis, these contributions come from the 
twist-3 local operators. It, however, also receives contribution from twist-2 local operators under 
the impulse approximation. Thus, the structure function $\mathsf{g}_{2}$ can be written as the sum 
of the twist pieces 
\begin{equation}
\mathsf{g}_{2} (x,Q^{2})=\mathsf{g}_{2}^{WW}(x,Q^{2})+\overline{\mathsf{g}}_{2} (x,Q^{2}),
\label{g2_evol}
\end{equation}   
where the twist-2 piece is given as \cite{Wandzura:1977qf}
\begin{equation}
\mathsf{g}_{2}^{WW} (x,Q^{2})=-\mathsf{g}_{1}(x,Q^{2}) + \int_{0}^{1} \frac{1}{y} 
\mathsf{g}_{1} (y,Q^{2}) \mathrm{d} y\,,
\end{equation}
while that of the twist-3 piece is
\begin{equation}
\overline{\mathsf{g}}_{2} (x,Q^{2})=\mathsf{g}_{1}(x,Q^{2})+\mathsf{g}_{2}(x,Q^{2}) 
-\int_{0}^{1} \frac{1}{y}\mathsf{g}_{1} (y,Q^{2}) \mathrm{d} y\,.
\end{equation}
The twist-2 part undergoes the ordinary evolution as in Eq.~(\ref{app:eveqn})
and the twist-3 piece is first parameterized by its moments
\begin{equation}
M_j(\mu^2)=\int dx\, x^{j-1} \overline{\mathsf{g}}_{2} (x,\mu^2)
\label{eq:momentsg2}
\end{equation}
that scale as \cite{Jaffe:1989xx}
\begin{equation}
M_j(Q^2)=\left[\frac{\ln(\mu^2)}{\ln(Q^2)}\right]^{\frac{\gamma_{j-1}}{\beta_0}}
\qquad {\rm with}\qquad
\gamma_{j-1}=2N_c\left[\psi(j)+\frac{1}{2j}+\gamma_E-\frac{1}{4}\right]\,.
\end{equation}
Here $\psi(j)$ is the logarithmic derivative of the $\Gamma$-function.
Then $\overline{\mathsf{g}}_{2} (x,Q^2)$ is obtained by expressing it in terms
of the evolved moments, {\it i.e.} by inverting Eq.~(\ref{eq:momentsg2}).

\enlargethispage*{0.5cm}


\end{document}